\documentclass[usenatbib]{mnras}
\usepackage{aas_macros}
\usepackage{amsmath}
\usepackage{amssymb}
\usepackage{color}
\usepackage{epsfig}
\usepackage{float}
\usepackage{graphicx}
\usepackage{latexsym}
\usepackage{morefloats}
\usepackage{natbib}
\usepackage{subfigure,multirow}
\usepackage{times}
\usepackage{soul,xcolor}
\setstcolor{red}
\usepackage[normalem]{ulem}
\newcommand{\plotone}[1]{\resizebox{0.95\hsize}{!}{\includegraphics{#1}}}

\newcommand{\plotthree}[3]{\center {\resizebox{0.99\hsize}{!}
{\includegraphics{#1}\hspace{0.3cm}\includegraphics{#2}\hspace{0.3cm}\includegraphics{#3}}}}

\newcommand{\Msun}{{~\rm M_\odot}}

\newcommand{\pc}{~\rm pc}
\newcommand{\kpc}{~\rm kpc}

\def\gsim { \lower .75ex \hbox{$\sim$} \llap{\raise .27ex \hbox{$>$}}}
\def\lsim { \lower .75ex \hbox{$\sim$} \llap{\raise .27ex \hbox{$<$}}}
\newcommand{\eagle}{\textsc{EAGLE}}
\newcommand{\gadget}{\textsc{Gadget}}
\newcommand{\emosaic}{\textsc{E-MOSAICS}}
\newcommand{\mosaic}{\textsc{MOSAICS}}

\newcommand{\refsec}[1]{Section~\ref{#1}}
\newcommand{\reftab}[1]{Table~\ref{#1}}
\newcommand{\reffig}[1]{Fig.~\ref{#1}}

\definecolor{blue-violet}{rgb}{0.54, 0.17, 0.89}
\definecolor{purple}{rgb}{0.5, 0., 0.5}

\definecolor{purple}{rgb}{0.5, 0., 0.5}

\title[Fornax GCs in E-MOSAICS]
{The survival of globular clusters in a cuspy Fornax}
\author[Shao et al.]
{\parbox{\textwidth}{
    Shi Shao$^{1}$\thanks{E-mail: shi.shao@durham.ac.uk}, 
Marius Cautun$^{2,1}$,
Carlos S. Frenk$^{1}$,
Marta Reina-Campos$^{3,4,5}$,
Alis J. Deason$^{1}$,
Robert A. Crain$^{6}$,
J. M. Diederik Kruijssen$^{3}$ and
Joel Pfeffer$^{6}$
\vspace{.20cm}}\\
$^1$ Institute for Computational Cosmology, Department of Physics, Durham University, South Road Durham DH1 3LE, UK \\
$^2$ Leiden Observatory, Leiden University, PO Box 9513, NL-2300 RA Leiden, the Netherlands \\
$^3$ Astronomisches Rechen-Institut, Zentrum f\"ur Astronomie der Universit\"{a}t Heidelberg, M\"onchhofstrasse 12-14, D-69120 Heidelberg, Germany\\
$^4$ Department of Physics \& Astronomy, McMaster University, 1280 Main Street West, Hamilton, ON, L8S 4M1, Canada \\
$^5$ Canadian Institute for Theoretical Astrophysics (CITA), University of Toronto, 60 St George St, Toronto, ON M5S 3H8, Canada \\
$^6$ Astrophysics Research Institute, Liverpool John Moores University, 146 Brownlow Hill, Liverpool L3 5RF, UK\\
}

\pubyear{2020}

\begin{document}
\label{firstpage}
\pagerange{\pageref{firstpage}--\pageref{lastpage}}
\maketitle

\begin{abstract}
  It has long been argued that the radial distribution of globular
  clusters (GCs) in the Fornax dwarf galaxy requires its dark matter
  halo to have a core of size $\sim 1$~kpc. We revisit this argument
  by investigating analogues of Fornax formed in \emosaic{}, a
  cosmological hydrodynamical simulation that self-consistently
  follows the formation and evolution of GCs in the \eagle{} galaxy
  formation model. In EAGLE, Fornax-mass haloes are cuspy and well
  described by the Navarro-Frenk-White profile. We post-process the
  \emosaic{} to account for GC orbital decay by dynamical
  friction, which is not included in the original model. Dynamical
  friction causes $33~$per cent of GCs with masses
  $M_{\rm GC}\geq4\times10^4\Msun$ to sink to the centre of their host
  where they are tidally disrupted. Fornax has a total of five GCs, an
  exceptionally large number compared to other galaxies of similar
  stellar mass. In the simulations, we find that only 3~per cent of
  the Fornax analogues have five or more GCs, while 30 per cent have
  only one and 35 per cent have none. We find that GC systems in
  satellites are more centrally concentrated than in field dwarfs, and
  that those formed in situ (45 per cent) are more concentrated
  than those that were accreted. The survival probability of a GC
  increases rapidly with the radial distance at which it formed
  ($r_{\rm init}$): it is 37 per cent for GCs with
  $r_{\rm init} \leq 1 \kpc$ and 92 per cent for GCs with
  $r_{\rm init} \geq 1\kpc$. The present-day radial distribution of
  GCs in \emosaic{} turns out to be indistinguishable from that in
  Fornax, demonstrating that, contrary to claims in the literature,
  the presence of five GCs in the central kiloparsec of Fornax does
  not exclude a cuspy DM halo.
\end{abstract} 

\section{Introduction}

One of the fundamental predictions of the standard cosmological model
($\Lambda$CDM) is that the density profiles of dark matter (DM) haloes
have cuspy profiles of the Navarro-Frenk-White (NFW) form, where the
inner DM density profile follows $\rho \propto r^{-1}$
\citep{Navarro1996,Navarro1997}. In contrast, some studies of stellar
motions and rotation curves of faint galaxies often appear to indicate
constant-density profiles at the centre, $\rho \propto r^0$
\citep[e.g.][]{Moore1994, Flores1994,Battaglia2008,Walker2011}. The
apparent discrepancy between theoretical predictions and observations
has become known as the core-cusp problem and several hypotheses have
been proposed to explain it. These include modifications of the
assumed nature of DM
\citep[e.g.][]{Spergel2000,Rocha2013,Shao2013,Kaplinghat2016,Schneider2017},
galaxy formation processes that alter the structure of the halo,
\citep[e.g.][]{Navarro1996b,Read2005,Pontzen2012,DiCintio2014,Benitez-Llambay2019},
and systematic biases in the interpretation of the observational data
\citep[e.g.][]{Oman2019}.

Dwarf galaxies are important cosmological probes since their inner
structure is particularly sensitive to the nature of DM; this has
served to stimulate numerous studies of the rotation curve and
kinematics of dwarfs. The kinematics of stars are tracers of the
potential and multiple analysis methods have been employed to infer
the mass profile of the DM halo in which they orbit, such as Jeans,
and more complex, modelling
\citep[e.g.][]{Gilmore2007,Walker2009,Amorisco2013,Pascale2018,Read2019}. However,
due to the relatively low number of stars with precise radial
velocities measurements and the lack of accurate proper motion
measurements, the current data are unable to distinguish robustly
between halo profiles with central cores or cusps in dwarfs
\citep[e.g.][]{Strigari2010,Strigari2017,Genina2018a,Genina2019b}.
Dwarfs contains not only stars but many of them also host globular
clusters (GCs)
\citep[e.g.][]{Mackey2003,Law2010,McConnachie2012,Forbes2018}, and
these can also be used to constrain the profile of DM haloes.
 
Even though the scatter in the number of GCs associated with dwarf
galaxies is large \citep[e.g.][]{Forbes2018}, their presence in dwarf
spheroidals like Fornax has long been considered a useful probe of the
DM distribution in these galaxies. In particular, the GC system in
Fornax has received considerable attention. It consists of five GCs
that are found at an average projected distance of $\sim1.1\kpc$ from
the centre \citep{Mackey2003}. They are typically very old,
with a lookback formation time, $t_{\rm age}\approx12$ Gyrs (Fornax 4
is an exception with $t_{\rm age}\approx 10$ Gyrs; see Table~1). The
interest in the Fornax GCs stems from comparing their ages to the
orbital decay times (the time required for the GC to sink to the
centre) due to dynamical friction. If the GCs formed at their
present-day positions, then the orbital decay time in a cuspy DM halo
would be shorter than their present age \citep[e.g.][]{Hernandez1998,
  Goerdt2006}; this discrepancy has been called the ``GC timing
problem''. One simple solution is to increase the decay timescales by
assuming larger initial radii for the GCs rather than assuming that they
formed at their present-day positions
\citep[e.g.][]{Angus2009,Boldrini2019, Meadows2020}. However, the
initial positions where the GCs were born are difficult to constrain
due to the long timescales involved.

An alternative solution is to assume that the inner density profile of
the Fornax halo has a kiloparsec-sized core. \citet{Goerdt2006} used
idealized N-body simulations to investigate the orbits of GCs under
different assumptions for the DM profile. They argued that when
dynamical friction is taken into account, the GCs would sink in a halo
with a flat inner density profile and stall at some radius for a few
gigayears. In contrast, in a cuspy halo, the GCs would gradually sink
all the way to the centre and be disrupted quickly by the central
tidal field or perhaps end up as a nuclear star cluster
\citep[e.g.][]{Tremaine1975}. \citet{Cole2012} also reported that if
the core is relatively large, GCs could be pushed out by the so-called
``dynamical buoyancy'' mechanism.
%\Shao{Another possible scenario is that a past dwarf-dwarf merger
%  expanded the orbits of the GCs \citep{Leung2020}}. 
Based on the argument that GCs would stall in a halo with a central
core, several studies have estimated the core radius of Fornax from
the present-day positions of its GCs \citep[e.g.][]{Read2006,
  Inoue2009, Cole2012, Petts2015, Kaur2018, Boldrini2019}. 

Another possibility has been proposed by \citet{Boldrini2020} who have
shown that if the GCs are surrounded by their own DM minihaloes, then
there is no ``GC timing problem'' even in cuspy haloes. Finally, it
is worth mentioning that tidal streams produced by the disruption of
GCs stripped from dwarf galaxies can perhaps be used to distinguish
between a central cusp or a core in the halo of the dwarf from which 
the GC was stripped \citep{Malhan2020}.

In this paper, we investigate if the present-day number and radial
distribution of the GCs in the Fornax dSph are consistent with
$\Lambda$CDM predictions in a cuspy DM halo. For this, we use one of
the \emosaic{} simulations that self-consistently model galaxy and GC
formation and evolution in a fully cosmological context
\citep{Pfeffer2018,Kruijssen2019a}. The presence of GCs does not
affect the DM halo profile and so the \emosaic{} dwarfs have the same
DM profiles as in the \eagle{} simulation, namely they are cuspy and
well fitted by the NFW parametrisation
\citep[e.g.][]{Schaller2015a,Benitez-Llambay2019,Bose2019}.

We proceed by selecting a sample of Fornax analogues in \emosaic{},
that is galaxies with similar stellar mass to the Fornax dSph, and
identify the GCs associated to each analogue. We cannot directly use
the number and positions of those GCs since \emosaic{} does not
include an accurate calculation of GC orbital decay due to dynamical
friction. Instead we account for this effect in
post-processing. Firstly we trace back the GCs to either their
formation time (for in-situ GCs) or their accretion time (for accreted
GCs). Then, starting from these initial positions and velocities, we
follow the orbital decay of the GCs by fitting an NFW profile to the
host halo and analytically integrating the orbits subject to dynamical
friction down to the present day. We account for the tidal disruption
of the GCs that sink to the centre of the host galaxy. Finally, we
compare the resulting present-day population of GCs in the simulation
against the distribution of GCs in the Fornax dSph.

This paper is organised as follows. In \refsec{sect:simul} we
describe the \emosaic{} simulations and, in
\refsec{sec:sample_selection}, we introduce our Fornax analogue sample
and examine the properties of their DM haloes and GCs. In
\refsec{sec:GC_orbit_evolution} we describe our method 
for modelling dynamical friction and tidal disruption of the GCs. In
\refsec{sect:result} we present our results on the number and radial
distribution of GCs in Fornax-mass dwarfs. We conclude with a
discussion of our results in \refsec{sec:discussion} and a short
summary of our main findings in \refsec{sect:conclusions}. 

\begin{table}
    \centering
    \caption{Selected properties of the Fornax GCs. The columns give:
      GC designation, projected distance from the centre of the Fornax
      dSph, stellar mass, age and metallicity. The projected distances
      are taken from \citet{Mackey2003} and updated for the current
      distance of the Fornax dSph of $147\kpc$
      \citep{McConnachie2012}. The remaining properties are taken from
      \citet{deBoer2016}.} 
    \begin{tabular}{ @{} lcccc @{} } 
        \hline\hline
        Name & $r_{p} \;[\rm{kpc}]$ & $M_{\rm GC} \;[10^5\; M_\odot]$ & Age [Gyrs] & [Fe/H]\\
        \hline 
        Fornax 1 & 1.72 & 0.42$\pm$0.10 & 12.1$\pm$0.8 & -2.5$\pm$0.3 \\ 
        Fornax 2 & 1.13 & 1.54$\pm$0.28 & 12.2$\pm$1.0 & -2.5$\pm$0.3 \\ 
        Fornax 3 & 0.46 & 4.98$\pm$0.84 & 12.3$\pm$1.4 & -2.5$\pm$0.2 \\ 
        Fornax 4 & 0.26 & 0.76$\pm$0.15 & 10.2$\pm$1.2 & -1.2$\pm$0.2 \\ 
        Fornax 5 & 1.54 & 1.86$\pm$0.24 & 11.5$\pm$1.5 & -1.7$\pm$0.3 \\
        \hline
        Fornax 6$^\dagger$ & 0.030 & $\sim0.29$ & -- & -- \\
        \hline\hline
        \\[-.2cm]
        \multicolumn{5}{ @{} p{\columnwidth} @{} }{
        $^\dagger$ This is a recently discovered faint and highly elliptical GC, which is possibly undergoing tidal disruption \citep{Wang.M.Y.2019}. We estimated its mass by assuming that it has the same mass-to-light ratio as Fornax 1, which is the Fornax GC with the closest V-band magnitude to Fornax 6.
        }
    \end{tabular}
    \label{tab:Fornax_GC_properties}
\end{table}

\section{Simulations}
\label{sect:simul}
The \emosaic{} (MOdelling Star cluster population Assembly In
Cosmological Simulations within \eagle{}) suite of cosmological
hydrodynamical simulations is an extension of the ‘Evolution and
Assembly of GaLaxies and their Environments’ (EAGLE) simulations
\citep{Crain2015,Schaye2015} that includes a subgrid model of stellar
cluster formation, evolution, and disruption
\citep{Kruijssen2011,Pfeffer2018}. For a detailed description of the
physical ingredients of the model, we refer the reader to
\citep{Pfeffer2018,Kruijssen2019a}.

Currently, the \emosaic{} project consists of two groups of
simulations. The first is a suite of 25 cosmological ``zoom''
simulations of MW-mass haloes
\citep{Pfeffer2018,Kruijssen2019a}. These contain only a small number
of Fornax analogues and are not used in this analysis. The second
group, which is the one we use, is a simulation of a full cosmological
volume in a periodic cube of side-length 34.4 (comoving)~Mpc (Crain et al.\ in prep.). The
volume of the simulation is 2.6 times larger than the \eagle{}
``high-resolution'' simulation which has the same resolution (labelled
Recal-L025N0752). The simulation follows the evolution of $1034^3$ DM
particles and an initially equal number of gas particles. The DM
particle mass is $1.2\times 10^6 \Msun$, and the initial gas particle
mass is $2.3\times 10^5 \Msun$. \emosaic{} assumes the
\textit{Planck-1} cosmology \citep{Planck2014} with cosmological
parameters:
$\Omega_{\rm m}=0.307, \Omega_{\rm b}=0.04825,
\Omega_\Lambda=0.693,h=0.6777,\sigma_8=0.8288$
and $n_{\rm s}=0.9611$, which are those used by the \eagle{}
project.

As in the \eagle{} project, the simulation we analyze here was
performed with a modified version of the \gadget{} code
\citep{Springel2005}, which includes state-of-the-art smoothed
particle hydrodynamics
\citep{DallaVecchia2012,Hopkins2013,Schaller2015a} and subgrid models,
such as element-by-element gas cooling, star formation, metal
production, stellar winds, and stellar and black hole feedback
\citep{Wiersma2009a,Springel2005a,Booth2009,Schaye2015}. The
parameters were calibrated so as to reproduce three present-day
observables: the stellar mass function, the galaxy size--mass
relation and the normalization of the relation between the masses of
supermassive black holes and the stellar mass of their host
galaxies. For a more detailed description we refer the reader to
\citet{Schaye2015}.

The semi-analytic model for \mosaic{} was coupled with \eagle{} to
track the formation and evolution of star clusters. The model is
calculated 
on-the-fly within \eagle{} since the time resolution required to
resolve the rapid change of the tidal field experienced by the GCs ($<1$ Myr),
which drives most of their disruption, is much finer than
the time interval between simulation snapshots (roughly  $70$
Myrs). The formation of stellar particles in the simulation
triggers the formation of a subgrid population of star clusters 
which inherit properties of their host stellar particles such as
position, velocity, age, and metallicity. 

The number of star clusters and their masses at formation are
determined by two parameters, the cluster formation efficiency (CFE,
\citealt{Bastian2008}) and an upper truncation mass scale in the
\citet{Schechter1976} initial cluster mass function ($M_{c,\star}$),
with a power-law index of -2 at low masses. These parameters are
described in terms of the local natal properties of the GC, such as
local ambient gas properties (pressure, density and mass) and stellar
velocity dispersion. Environments with higher gas pressure lead to the
formation of more star clusters which can also be more massive
\citep{Kruijssen2012,Reina-Campos2017}.

Four GC models are applied to the simulation in parallel. The fiducial
model allows the CFE and $M_{c,\star}$ to vary as a function of
properties of the local environment in which the stars are formed
\citep{Kruijssen2012,Reina-Campos2017}. The other three keep either,
or both of the CFE and $M_{c,\star}$ fixed. For more details of the
cluster formation models, we refer the reader to
\citet{Pfeffer2018} and \citet{Reina-Campos2019}. After the clusters form, they lose mass
due to stellar evolution (according to the fractional mass-loss of the
parent stellar particle) and by dynamical effects such as two-body
relaxation and tidal shocks which are based on the strength and change
of the local tidal field respectively \citep{Kruijssen2011}.
%The GCs below a certain mass ($10^2\Msun$) or those fail the post-processing dynamical friction selection criterion are removed from the snapshots \citep{Pfeffer2018}. \MC{I don't think we need this last sentence.}

The \emosaic{} simulations have been able to reproduce a broad range
of observational properties such as the deficit of massive metal-poor
GCs (i.e.~the ``blue tilt") in galaxies across a wide range of
environments \citep{Usher2018} and the diversity of age-metallicity
relations of GCs in different galaxies \citep{Kruijssen2019a}. The
$M_{\rm GC}/M_{\rm halo}$ and $M_{\rm GC}/M_{\star}$ relations
predicted by the simulation are also in good agreement with
observations \citep{Bastian2020}. Additionally, the large sample of
MW-mass galaxies allows the assembly history of our MW and the GC
formation history to be probed by reference to the observed
present-day GC populations
\citep{Kruijssen2019b,Kruijssen2020,Reina-Campos2019,Pfeffer2020,TrujilloGomez2020}. The simulations
also have strong implications for the origin of the stellar bulge and
stellar halo whose masses are made up, in part, of GC remnants
\citep{Hughes2020,Reina-Campos2020}. In addition, the simulations can
be used to make predictions for the properties of the recently
observed GCs in the M31 system \citep{Hughes2019}.

The halo and galaxy catalogues in \emosaic{} have been built using the
tools described by \citet{Schaye2015}. Haloes are initially identified
by the friends-of-friends (FOF) algorithm \citep{Davis1985} with a
linking length $0.2$ times the mean interparticle separation. The
resulting FOF groups are further split into gravitationally bound
substructures using the \textsc{subfind} code
\citep{Springel2001,Dolag2009}, applied to the total matter
distribution (dark matter, gas and stars) associated with each FOF
group. The central subhalo is defined as the subhalo that contains the
most bound particle, while the remaining subhaloes are classified as
satellites. The stellar distribution associated with the main subhalo
is identified as the central galaxy. The central haloes are
characterized by the mass, $M_{200}$, and radius, $R_{200}$, that
define an enclosed spherical overdensity of $200$ times the critical
density. The position of each galaxy, for both centrals and
satellites, is given by their most bound particle.

\begin{figure}
	\plotone{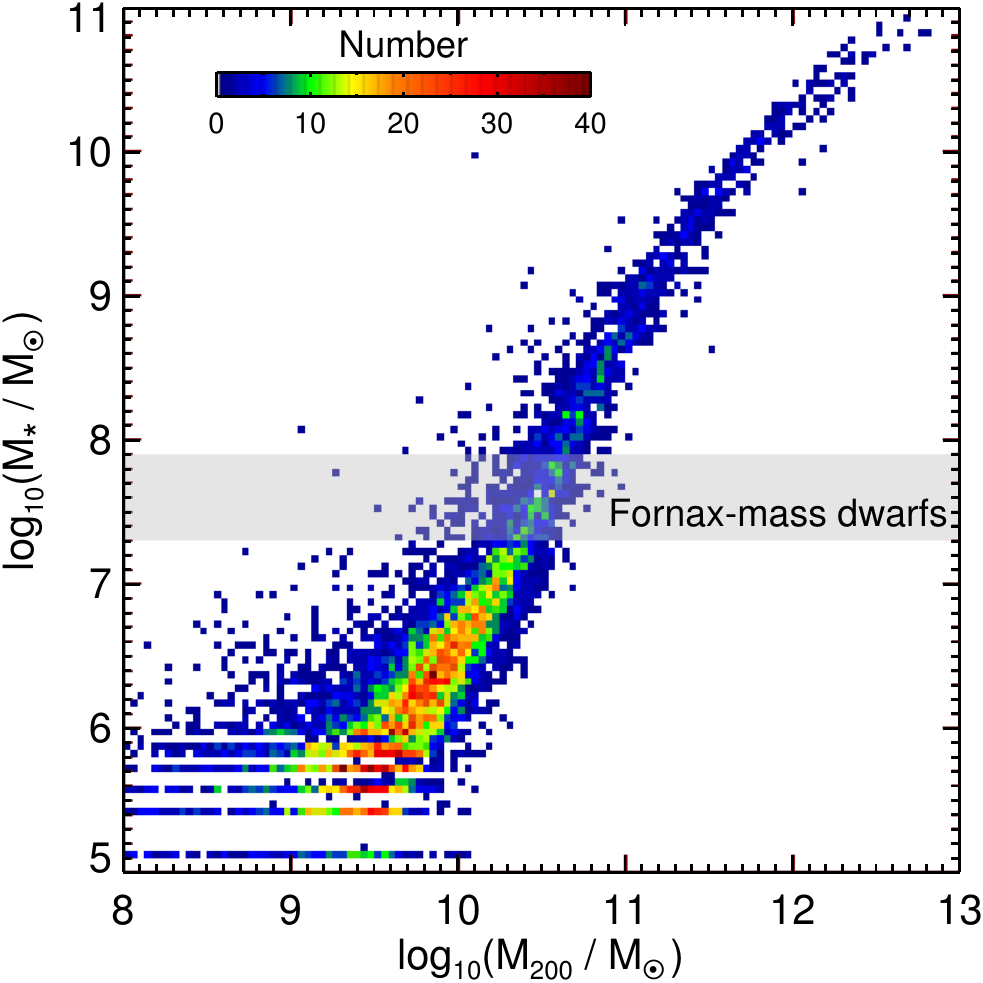}
	\caption{ The relation between stellar mass, $M_{\star}$, and
		total halo mass, $M_{200}$, for central galaxies in the
		simulation. The colours indicate the number of
		galaxies in each halo and stellar mass bin (see legend). The
		grey shaded region shows galaxies with stellar masses in the
		range $2-8\times 10^{7} \Msun$, which corresponds to our
		sample of field Fornax-mass dwarfs (we also select Fornax-mass
		satellites, which are not shown in this diagram).}
	\label{fig:Fornax-sample}
\end{figure}

\reffig{fig:Fornax-sample} shows the relation between the stellar mass
of central galaxies and the mass of their host haloes in
\emosaic{}. The satellite galaxies are not shown since the total mass
associated with the subhalo changes as it orbits in the main halo. The
scatter in the stellar-to-halo mass relation for Fornax-mass dwarfs is
larger than for more massive galaxies, but is significantly smaller
than for lower mass dwarfs \citep{Sawala2015}.

%%%%%%%%%%%%%%%%%%%%%%%%%%%%%%%%%%%%%%%%%%%%%%%%%%%%%%%%%%%%%%%%%%%%%
%%%%%%%%%%%%%%%%%%%%%%%%%%%%%%%%%%%%%%%%%%%%%%%%%%%%%%%%%%%%%%%%%%%%%

\section{Sample selection and Methods}
\label{sec:sample_selection}

%%%%%%%%%%%%%%%%%%%%%%%%%%%%%%%%%%%%%%%%%%%%%%%%%%%%%%%%%%%%%%%%%%%%%
\begin{figure*}
	\plotthree{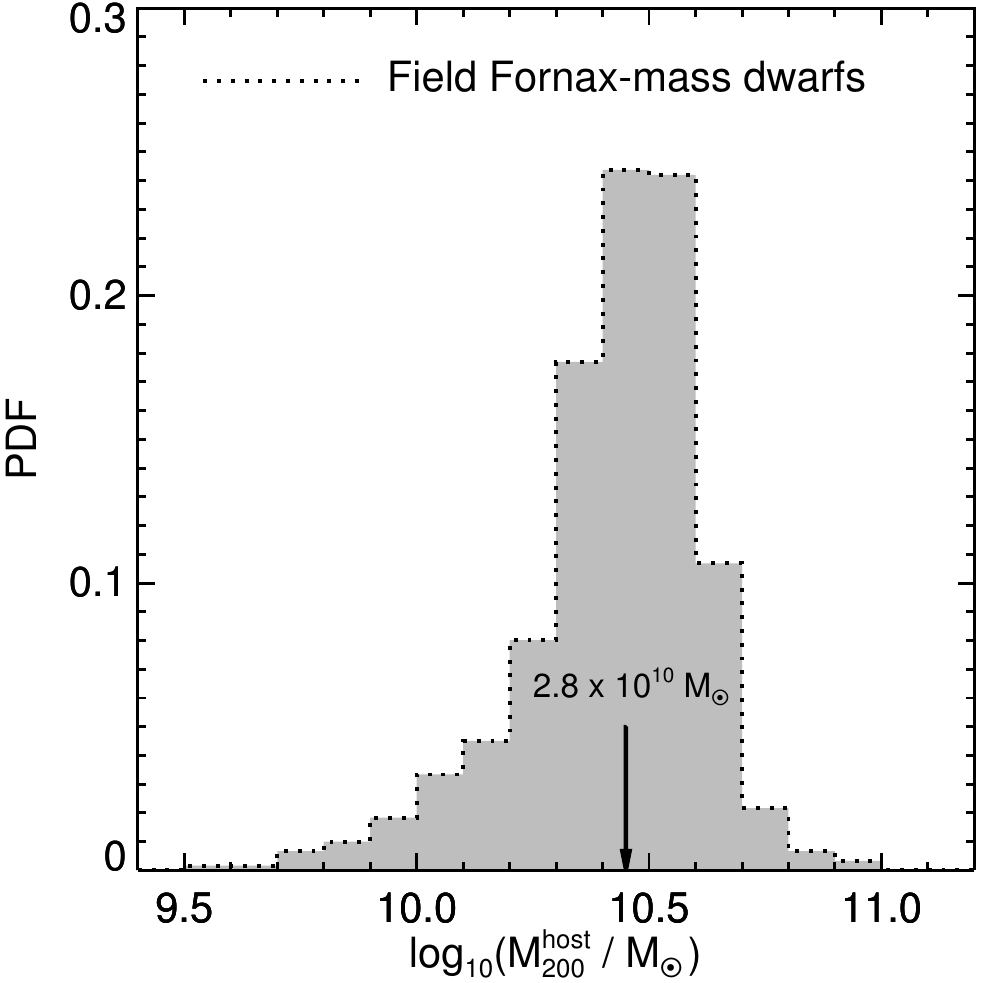}{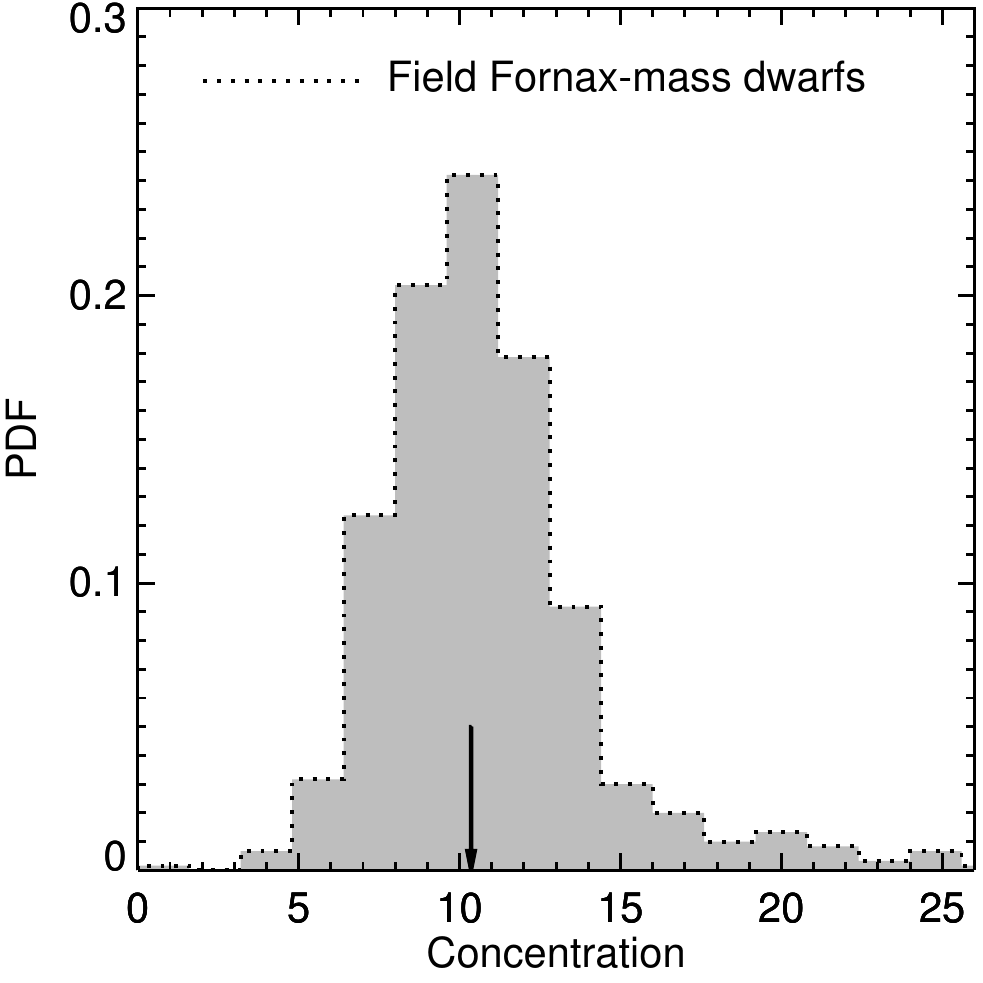}{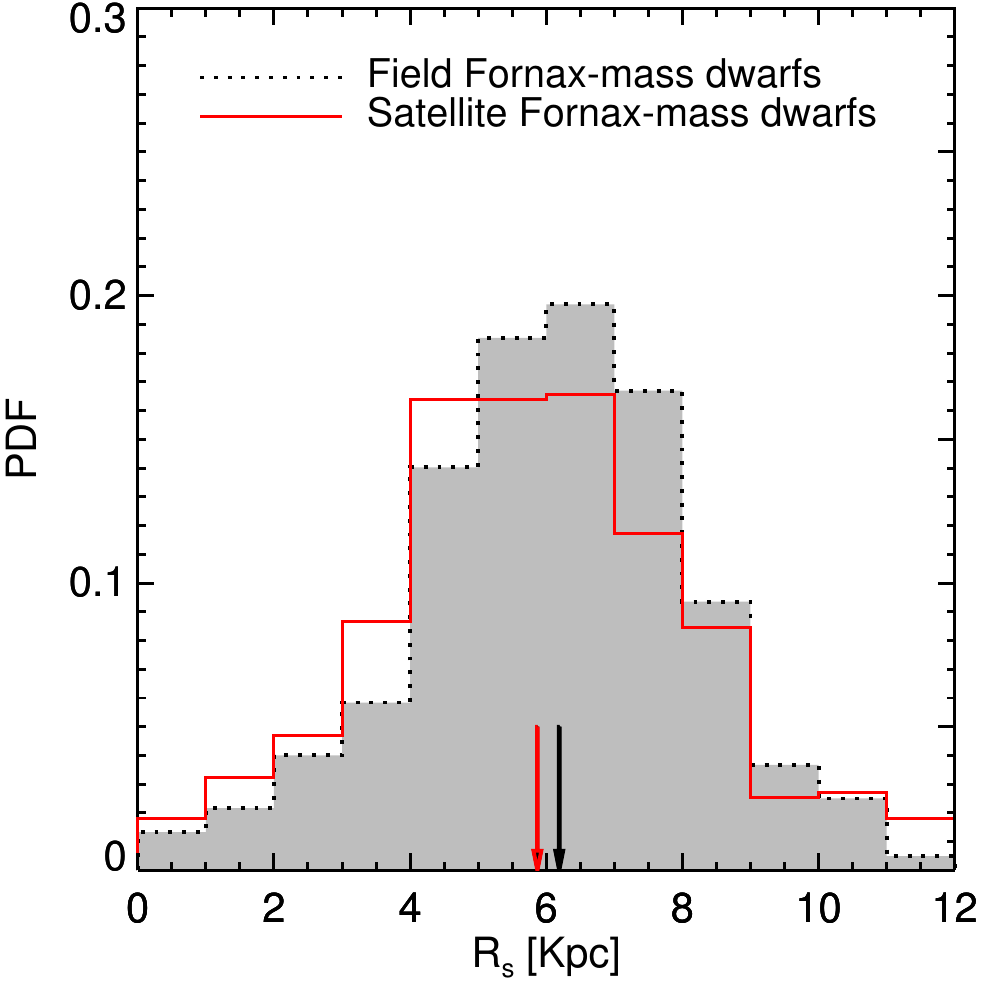}
	\caption{ The properties of the DM haloes that host a
          Fornax-mass dwarf galaxy. The plot shows the PDF of the halo
          mass (left panel), concentration (centre panel), and scale
          radius (right panel). The results are for present-day field
          galaxies, since the same properties for satellite galaxies
          are not directly comparable (see text for discussion on
          this). However, we do show the distribution of scale radii
          for satellites as the solid red line in the right panel. The
          vertical arrows indicate the median of each PDF. 
	}
	\label{fig:Pdf_Fornax}
\end{figure*}
%%%%%%%%%%%%%%%%%%%%%%%%%%%%%%%%%%%%%%%%%%%%%%%%%%%%%%%%%%%%%%%%%%%%%

\subsection{The Fornax analogue sample selection and their DM halo
  properties} 
We select
Fornax analogues, to which we also refer as Fornax-mass dwarfs, in the
simulation by requiring that they have a stellar mass in the range,
$M_\star \in [2, 8] \times 10^{7}\Msun$. This results in 1154
objects, of which 599 are field galaxies and 555 are satellites. The mass range used for the selection corresponds to a factor
of two variation around the Fornax dSph stellar mass, which we take to
be $4\times 10^{7}\Msun$ \citep{deBoer2012}. The mass range is
relatively wide since we need a large sample of Fornax-mass dwarfs for
good statistics and a factor of 2 is representative of the
  uncertainty in Fornax's mass estimates. A typical
Fornax-mass dwarf in \emosaic{} is resolved with ${\sim}2\times10^4$
DM particles, and hundreds of star particles, which allows for a
robust determination of its present day properties as well as its
formation history.

We study the general properties, such as halo mass, concentration, and
scale radius, $R_s$, of the Fornax-mass dwarf sample in
\reffig{fig:Pdf_Fornax}. The left panel shows the distribution of the
host halo mass, $M_{\rm 200}$, for the field Fornax-mass dwarfs. The
satellites are not shown since they are tidally stripped by the hosts
after their infall. Nevertheless, for satellite dwarfs, we refer to
the $M_{\rm 200}$ of their main progenitors, before they became a
satellite of a more massive host, as their host halo mass. The
distribution for the field sample peaks at a value of
$2.8^{+1.1}_{-1.0}\times 10^{10} \Msun$ (68~per cent confidence
limit), with sharp drop-offs on both sides; this is in agreement with
abundance matching predictions and results from other hydrodynamical
simulations \citep[e.g.][]{Moster2013,Sawala2015,Onorbe2015}. The
width of the distribution largely reflects the fact that haloes of a
given mass can have a range of different concentrations; higher
concentration haloes, which typically form early, have more time to
form stars and experience less efficient feedback.

The concentration and scale radius of the host
haloes are calculated by fitting the spherical DM density within
a radial distance of $50\kpc$ from the centre (roughly the median
$R_{\rm 200}$ of the field Fornax-mass haloes) to an NFW profile
\citep{Navarro1996,Navarro1997}:  
\begin{equation}
    \rho(r) = \dfrac{\rho_0}{\dfrac{r}{r_{\rm s}}\left(1+\dfrac{r}{r_{\rm s}}\right)^2}
    \label{eq:NFW_profile} \;,
\end{equation}
where $\rho_0$ and the scale radius, $r_{\rm s}$, are parameters (see appendix \ref{appendix:density_profile}). For
the satellites, we fit the DM profile of the main progenitor just before
infall. The concentration is given by the ratio $c=r_{\rm
  s}/R_{200}$. The distribution of concentration values is shown in
the middle panel of \reffig{fig:Pdf_Fornax}; it has a median value of
$c=10.5$ for the field Fornax-mass dwarfs. The result is in good
agreement with previous high-resolution cosmological simulations
\citep[e.g.][]{Hellwing2016}.

The right panel of \reffig{fig:Pdf_Fornax} shows the distribution of
the scale radius, $r_{\rm s}$, for both the field and satellite
samples. The field sample has a median value of $r_{\rm s}=6.2 \kpc$,
slightly larger than that of the satellite sample,
$r_{\rm s}=5.9\kpc$. The two samples have similar $z=0$ stellar mass
by construction; thus the difference in $r_{\rm s}$ between the two
samples is due to satellites having experienced tidal stripping and
also, potentially, to small differences in the assembly history of
satellite versus central galaxies.

\subsection{GCs sample selection and their formation time}

Here we compare the \emosaic{} predictions with the five most massive
GCs in the Fornax dSph. These have been known for a long time and have
been thoroughly studied \citep[e.g.][]{deBoer2016}; we give a few
selected properties of the Fornax GCs in Table
\ref{tab:Fornax_GC_properties}. More recently, \citet{Wang.M.Y.2019}
discovered a sixth GC in Fornax. It has a lower mass than the other
five and is possibly undergoing tidal disruption. For simplicity, and
better to compare with previous works, we limit our comparison to the
five well-known Fornax GCs.

To identify analogues of the Fornax GC population, we proceed by
selecting all GCs that at $z=0$ are associated with a Fornax analogue
and have a stellar mass above
$4\times10^4\Msun$.
The GC stellar mass threshold corresponds to the mass of the Fornax-1
GC, the least massive of the five Fornax GCs we study
here. Our selection results in 2133 GCs, out of which only 1439
survive to the present day (the others were tidally disrupted after
dynamical friction dragged them to the centre -- we describe this in
detail in \refsec{sec:model}).

GCs that are associated with their $z=0$ host may have formed inside
another galaxy that was subsequently accreted into the present-day
host. Such GCs can have different properties (e.g. age, metallicity,
radial distribution) from those born {\em in situ}. We study this by
splitting our sample into {\em in situ} and accreted GCs according to whether
they were formed in the main ({\em in situ}) or in a sub- (accreted) branch of
the progenitor of the $z=0$ host galaxy of each GC. We find that
around half ($45~$per cent) of the GCs formed {\em in situ}.

\begin{figure}
	\plotone{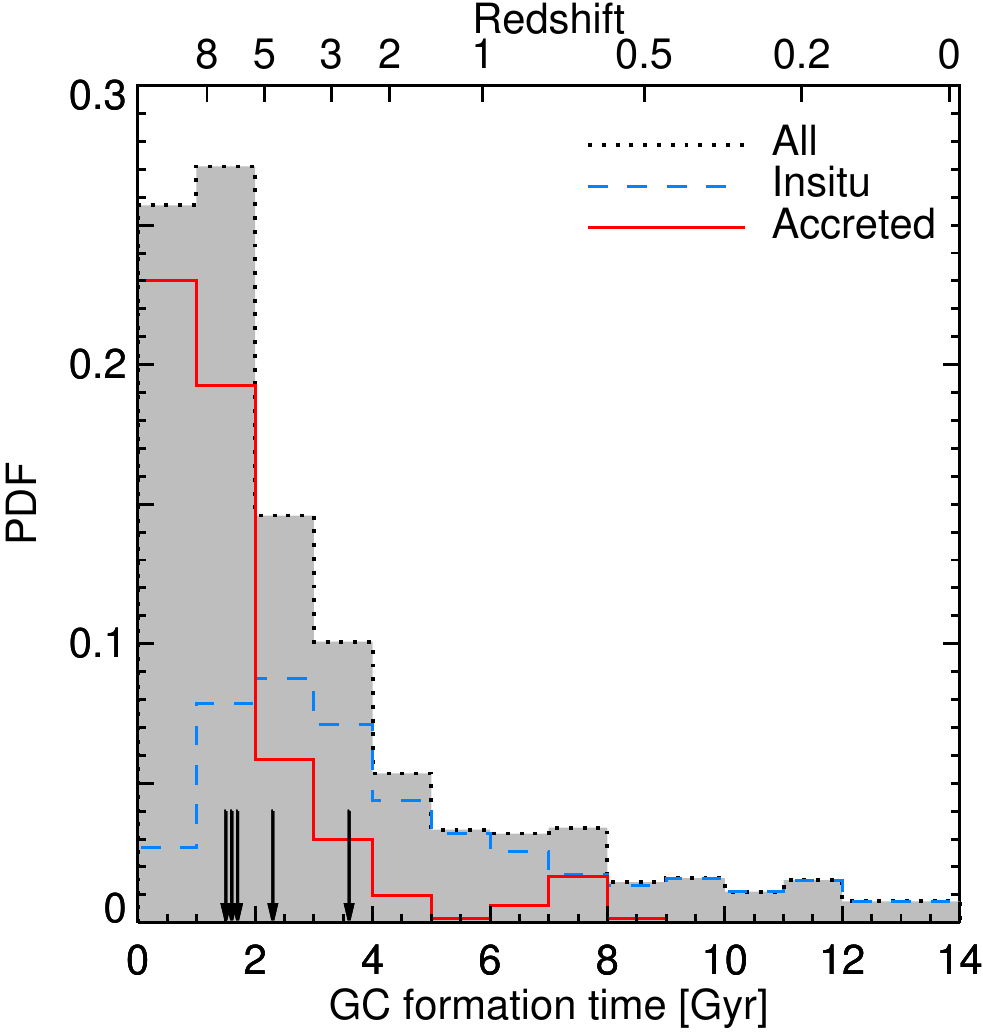}
	\caption{ The distribution of formation times for GCs found in
          Fornax-mass dwarfs at $z=0$. The plot shows the PDF for all
          GCs (black dotted line), {\em in situ} (blue dashed line)
          and accreted (red solid line) GCs. Arrows indicate the
          formation time of Fornax's GCs \citep{deBoer2016}. The
          left-hand side of the plot corresponds to the Big Bang and
          $t=13.8$ Gyrs to present day. }
	\label{fig:Pdf_GC_tform}
\end{figure}

\reffig{fig:Pdf_GC_tform} shows the probability distribution
function (PDF) of GC formation times for both {\em in situ} and
accreted objects that survive to $z=0$. Overall, GCs in \emosaic{}
have very early formation times, with more than half of the full
sample having formed within the first $2$ Gyrs after the Big Bang; this is
consistent with the inferred ages of the GCs in Fornax. The only exception is
Fornax 4 which is $\sim10$ Gyrs old and is relatively `younger' than
the other four GCs, as indicated by its more metal-rich stellar population
\citep{deBoer2016}. 

More interestingly, when comparing the accreted and {\em in situ} GC
populations, we find that $77$~per cent of accreted GCs formed early
($\sim12$ Gyrs old), while this fraction is only $23$~per cent for the
{\em in situ} ones. The variation in formation times between the two
GC populations is expected, since there are at least two processes
that act differently in the two populations. The accreted population
is {brought in} by mergers, and most such mergers take place at
$z\gtrsim 2$; {by construction}, those GCs must have formed before the
merger. Secondly, as we shall discuss later, {\em in situ} GCs are
more radially concentrated and thus dynamical friction is more
efficient at causing them migrate to the centre, where they are
tidally disrupted. This would suggest that recently formed {\em in
 situ} GCs are more likely to survive to $z=0$ than their older
siblings. Both of these processes lead to an excess of younger {\em in
 situ} GCs.

It is also worth noting that Fornax has two or more distinctive
populations of stars: a metal poor and a younger, more compact, metal
rich population \citep[e.g.][]{Battaglia2006}. One way to obtain such
configurations in simulations is through a major merger between two
dwarfs \citep[e.g.][]{Benitez-Llambay2016,Genina2019}, raising the
intriguing possibility that Fornax might have had one or more such
mergers \citep[see also][]{Yozin2012}.

%%%%%%%%%%%%%%%%%%%%%%%%%%%%%%%%%%%%%%%%%%%%%%%%%%%%%%%%%%%%%%%%%%%%%
%%%%%%%%%%%%%%%%%%%%%%%%%%%%%%%%%%%%%%%%%%%%%%%%%%%%%%%%%%%%%%%%%%%%%

\section{GC orbital evolution}
\label{sec:GC_orbit_evolution}

In the \emosaic{} model GCs are associated with stellar particles
whose dynamics they follow. Like the stars, GCs are collisionless
objects and thus this is a reasonable approximation, except for one
aspect: GCs can experience significant dynamical friction as they
orbit in their host galaxy. This process is central to this study and
thus we need to supplement the \emosaic{} model with a treatment of
dynamical friction. We do this by post-processing the simulated GC
populations from \emosaic{} as follows. We trace back each GC
associated with a present-day Fornax analogue to the simulation
snapshot closest to the time when the GC formed (for {\em in situ}
objects) or to the time when it was accreted into its $z=0$ host (for
accreted GCs). The corresponding positions and velocities, which are
given by \emosaic{}, are taken as the starting position of each GC and
its orbit is modelled in the potential of its host galaxy accounting
for dynamical friction. Finally, we follow the orbit of each GC until
the present day keeping track of whether it got close enough to the
galaxy centre to experience tidal disruption. In this section, we
present the model used to follow the GC orbit evolution.

\subsection{Dynamical friction}
\label{sec:model}
We follow the evolution of each GC separately in the potential of the
DM halo and stellar component of its host galaxy. The DM halo is
modelled as an analytic Navarro, Frenk \& White \citep[][hereafter 
NFW]{Navarro1996,Navarro1997} profile with total mass, $M_{200}$, and
concentration, $c$, whose potential is given by:
\begin{equation}
	\Phi_{\rm halo} = -\frac{G M_{200}}{r}
    	\frac{ \ln\left(1+ r/r_{\rm s}\right) }{\ln(1+c)-c/(1+c) }
        \label{eq:potential_NFW} \;,
\end{equation}
where $r_{\rm s}$ denotes the scale radius of the DM halo. The halo
parameters, $c$, $r_{\rm s}$ and $M_{200}$, are inferred by fitting an
NFW profile to the host DM halo at $z=0$ for field Fornax-mass
galaxies and at infall for the Fornax-mass galaxies that are
satellites at the present day. In the latter case, we assume that the
GC orbit is determined only by its Fornax-mass host, and we ignore
additional forces coming from the more massive system to which the
Fornax-mass satellite belongs. This approximation is valid as long as
the inner region, $r \lsim{} r_{\rm s}$, of the Fornax-mass satellite
does not undergo severe tidal disruption.

The GC dynamics are dominated by the DM halo potential, but, for
completeness, we also include the stellar potential. We model the
stellar component of the Fornax-mass galaxy as a Plummer profile
with potential:
\begin{equation}
	\Phi_{\rm stars} = -\frac{G M_{\star}}{\sqrt{r^2+b^2}}
        \label{eq:potential_Plummer} \;,
\end{equation}
where $M_{\star}$ denotes the stellar mass and the parameter, $b$, 
is given by $b=R_{1/2}/1.3$, with $R_{1/2}$ the 3D stellar half-mass
radius. The Plummer profile gives a good match to the stellar density
profile of Fornax and other dwarf spheroidal galaxies around the Milky
Way \citep[e.g. see][]{Wang2018z}.

We implement dynamical friction as a deceleration experienced by the
GC while orbiting within the host halo of its galaxy. We assume that
the deceleration is given by Chandrasekhar's formula, 
\begin{equation}
	\frac{{\rm d}{\bf v}}{{\rm d}t} = - \frac{ 4 \pi G^2 M_{\rm GC} \rho \ln\Lambda } {v^2} 
    	\left[ {\rm erf}(X) - \frac{2X}{\sqrt\pi} e^{-X^2} \right]\frac{{\bf v}}{v}
	\label{eq:chandra_dyn_fric} \;,
\end{equation}
\citep{Binney2008}, where $G$ is the gravitational constant,
$M_{\rm GC}$ is the GC mass, ${\bf v}$ is the relative velocity of the
GC with respect to its galaxy, $\rho$ is the density of the DM halo at
the GC's position, and $X = v/(\sqrt{2}\sigma_v)$, with $\sigma_v$ the
local 1D velocity dispersion of the DM halo. We take the Coulomb
logarithm as:
\begin{equation}
    \ln \Lambda = \ln \frac{b_{\rm{max}} \sigma_v^2}{G M_{\rm GC}}
    \label{eq:coloumb_logarithm} \;,
\end{equation}
\citep{Goerdt2006}, where $b_{\rm max}$ is the largest impact
parameter to be considered. By comparing with high resolution
simulations of GC orbit evolution, \citet{Goerdt2006} found that for
cuspy haloes, $b_{\rm max}=0.25\kpc{}$ is the best fitting value. We
include only the dynamical friction arising from the DM halo, which is
the dominant effect, and neglect any contribution from the stellar
distribution.
 
When integrating the orbits of the GCs we keep the potential of the DM
halo fixed at all times. This is a reasonable approximation since,
while the halo mass can grow by factors of several since the time most
GCs formed, ${\sim}12$ Gyrs ago, the growth takes place by adding new
mass to the outskirts of the halo while leaving the inner region
mostly unchanged \citep{Wang.Jie2011}. Most stars and GCs orbit in the
inner few kiloparsecs of the halo and thus their orbits will not be
affected by mass growth at the halo outskirts. However, this
approximation ceases to be valid when a halo undergoes major
mergers. In that case, the GC orbits can also be affected and, on
average, they are pushed slightly towards higher energy and more
extended orbits \citep[e.g.][]{Benitez-Llambay2016}. This would reduce
the effect of dynamical friction and slow down the orbital decay of
GCs. Thus, by not accounting for the effect of major mergers, we are
likely overestimating the number of GCs that sink to the centre of
their host.
%Nonetheless, this does not affect our conclusion, since, as we sha we find

\subsection{Tidal disruption of GCs}
To account for the disruption of GCs by the tidal field of their host
galaxy we calculate their tidal radius. For an NFW profile the tidal
radius for a GC on a circular orbit at distance, $r$, from the galaxy
centre is given by: 
\begin{equation}
    r_{\rm tidal} = (r+r_{\rm s}) \left( \frac{M_{\rm GC}}{M_{\rm
          200}} \frac{r}{3r+r_{\rm s}} \right)^\frac{1}{3}
    \label{eq:tidal_radius} \;,
\end{equation}
\citep{Renaud2011,Orkney2019}. This expression neglects the tides
arising from the stellar distribution of the host galaxy, which are
much smaller than the tidal field of the DM halo. We calculate the
tidal radius at each point along the GC's orbit and consider the
GC to be destroyed by the tidal field of its host when the tidal radius
is comparable to the half-mass radius of the GC. As GCs lose mass due
to two-body relaxation as well as tidal stripping, their half-mass
radius increases. According to the N-body simulations of
\citet{Orkney2019}, GCs evolving in cuspy profiles can reach a
half-mass radius of ${\sim}6\pc$, after which the mass loss rate
increases rapidly and the GC is disrupted shortly thereafter. Based on
these results, we assume that GCs are fully disrupted if at any point
along their orbit the tidal radius becomes smaller than $6\pc$.

\subsection{Decay of the GC orbits}
%%%%%%%%%%%%%%%%%%%%%%%%%%%%%%%%%%%%%%%%%%%%%%%%%%%%%%%%%%%%%%%%%%%%%
\begin{figure}
	\plotone{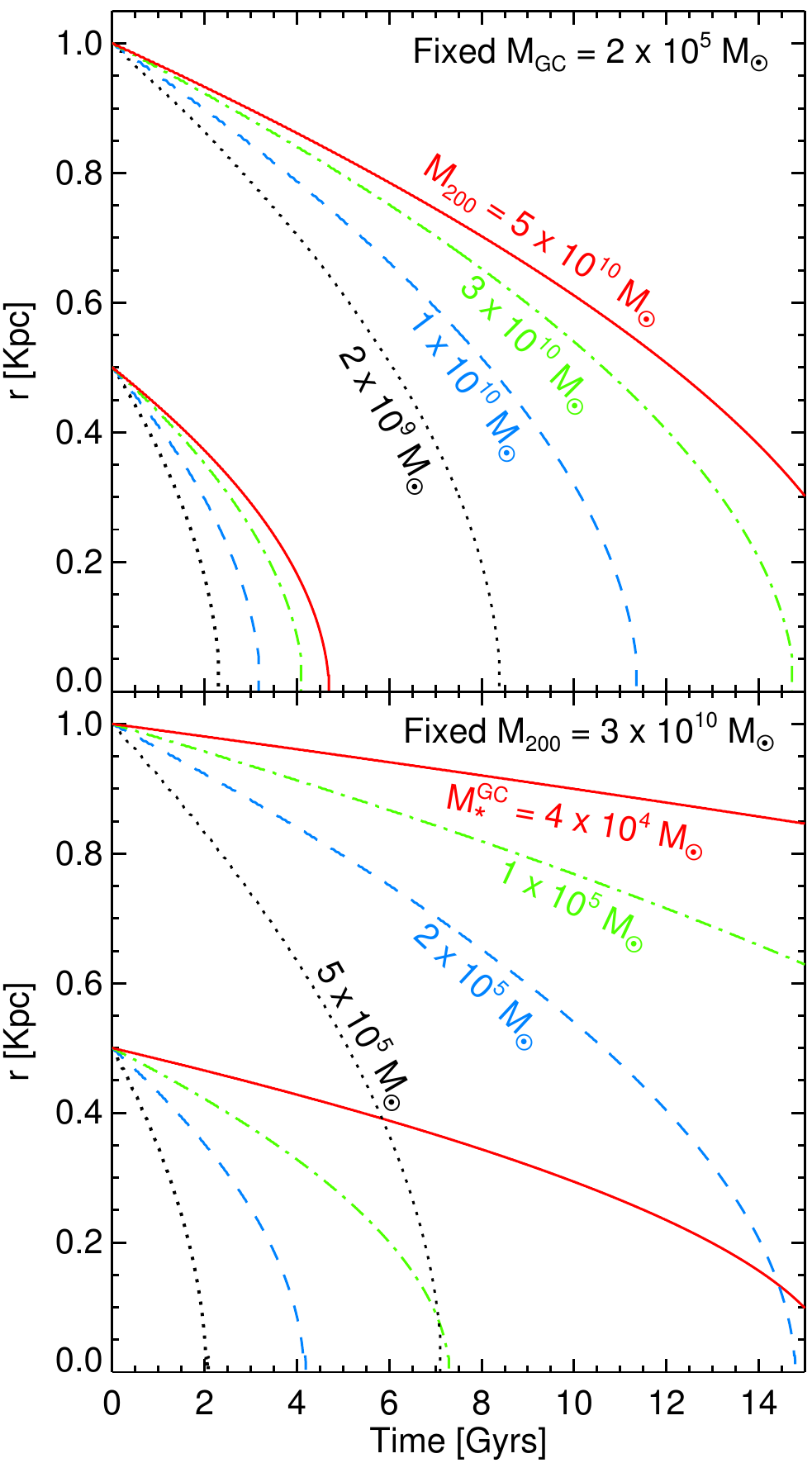}
	\caption{ The time evolution of the radial distance for two
          GCs on circular orbits whose starting positions are at
          $0.5\kpc$ and $1\kpc$ from the centre of its host. The top
          panel shows how the orbits change when varying the host halo
          mass (here, the GCs have the same mass, $M_{\rm
            GC}=2\times10^5\Msun$). The bottom panel shows how the
          orbits change when varying the GC mass (here, the host halo
          has the same mass, $M_{\rm 200}=3\times10^{10}\Msun$).} 
	\label{fig:GC_orbit_M200}
\end{figure}
%%%%%%%%%%%%%%%%%%%%%%%%%%%%%%%%%%%%%%%%%%%%%%%%%%%%%%%%%%%%%%%%%%%%%

The orbital decay time of a GC is mainly determined by two factors: 1)
the initial distance to the center of its host, and 2) the mass ratio
between the GC and the host halo. We illustrate this in
\reffig{fig:GC_orbit_M200}, where we show GC orbits for two starting
positions and a selection of GC and host halo masses. In all cases, we
start with GCs on circular orbits; in reality, \emosaic{} predicts
highly elliptical orbits even for the {\em in situ} GCs of our
Fornax-analogue sample, with a median ellipticity at birth of 0.64
(and 16 to 84 per centile of the ellipticity distribution of 0.36 and
0.84 respectively). However, here we want to illustrate the
systematic effect of dynamical friction, which is most clearly seen
for circular orbits.

In the first example, in the top panel in
\reffig{fig:GC_orbit_M200}, we fix the GC mass to $M_{\rm
  GC}=2\times10^5\Msun$, which is the median mass of Fornax's GCs, and
vary the mass of the host halo. We select a reference halo mass,
$M_{200}=3\times10^{10}\Msun$, which corresponds to the median halo
mass in our Fornax analogue sample, as well as a much lower mass,
$M_{200}=2\times10^{9}\Msun$, which was the value used by
\citet{Goerdt2006}. We also present orbits for host halo masses
$M_{200}=1$ and $5\times10^{10}\Msun$, which correspond to roughly the
10 and 90 per centile of the distribution of halo masses for our sample
of Fornax-mass dwarfs. 

We show the orbital evolution for two starting distances from the host
centre: $0.5\kpc$ and $1\kpc$. As expected, GCs that are initially
closer to the centre sink more quickly than those {which} start
further away. GCs with $r_{\rm init}=0.5\kpc$ sink to the center
within $\sim2$ Gyr in the least massive halo ($M_{\rm
  200}=2\times10^9\Msun$), and $\sim5$ Gyr in the most massive host halo
($M_{\rm 200}=5\times10^{10}\Msun$). The sinking time of GCs 
increases when increasing $r_{\rm init}$. For $r_{\rm init}=1\kpc$, the
GC can survive for a Hubble time in the two most massive hosts, but
not in the lower mass haloes. 

In view of the diversity of the mass of the Fornax GCs, in the bottom
panel we study the variation in orbits with GC mass. In this case, we
fix the halo mass to $M_{200}=3\times10^{10}\Msun$, the median value
for Fornax-mass hosts in the \emosaic{} simulation (see
\reffig{fig:Pdf_GC_tform}). We show orbits for four GC masses, from
$4$ to $50\times10^{4}\Msun$. The lowest mass GC experiences the
weakest orbital decay, and survives for a Hubble time even when
starting at $r_{\rm init}=0.5\kpc$. In contrast, the most massive GC
sinks rapidly to the centre, with a sinking time of $2$ and $7$ Gyrs
depending on whether it starts at 0.5 or $1\kpc$ from the centre.
This suggests that the most massive GC in Fornax, Fornax-3, must have
formed significantly farther than its present-day position ($0.46\kpc$
projected distance); we discuss this point in greater detail in
Sections~\ref{sec:discussion} and in \reffig{fig:future_past}.

The oldest GCs can lose up to 60~per cent of their initial mass due to
dynamical effects, such as two-body relaxation, tidal stripping and stellar evolution
(see Appendix~\ref{appendix:GC_mass_evolution}). We account for this
effect by having a time-varying mass for our GCs, with the mass
decreasing in time according to the prescription detailed in Appendix
\ref{appendix:GC_mass_evolution}.

\section{Results}
\label{sect:result}

We now compare the number and radial distribution of GCs in the
Fornax dSph with the \emosaic{} predictions. These predictions have
been post-processed to include dynamical friction and exclude GCs that
were tidally disrupted after sinking to the centre of their host, as
described in the previous section. 

%%%%%%%%%%%%%%%%%%%%%%%%%%%%%%%%%%%%%%%%%%%%%%%%%%%%%%%%%%%%%%%%%%%%%
\begin{figure}
	\plotone{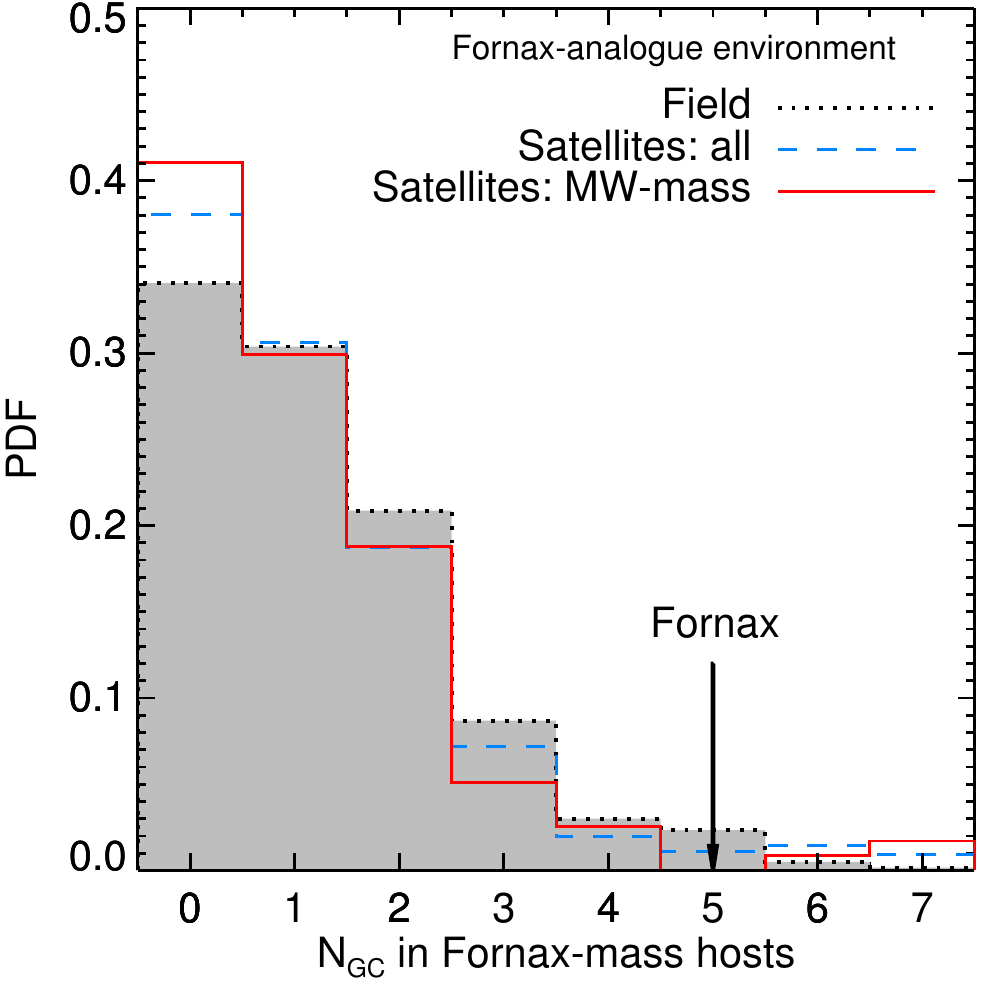}
    \plotone{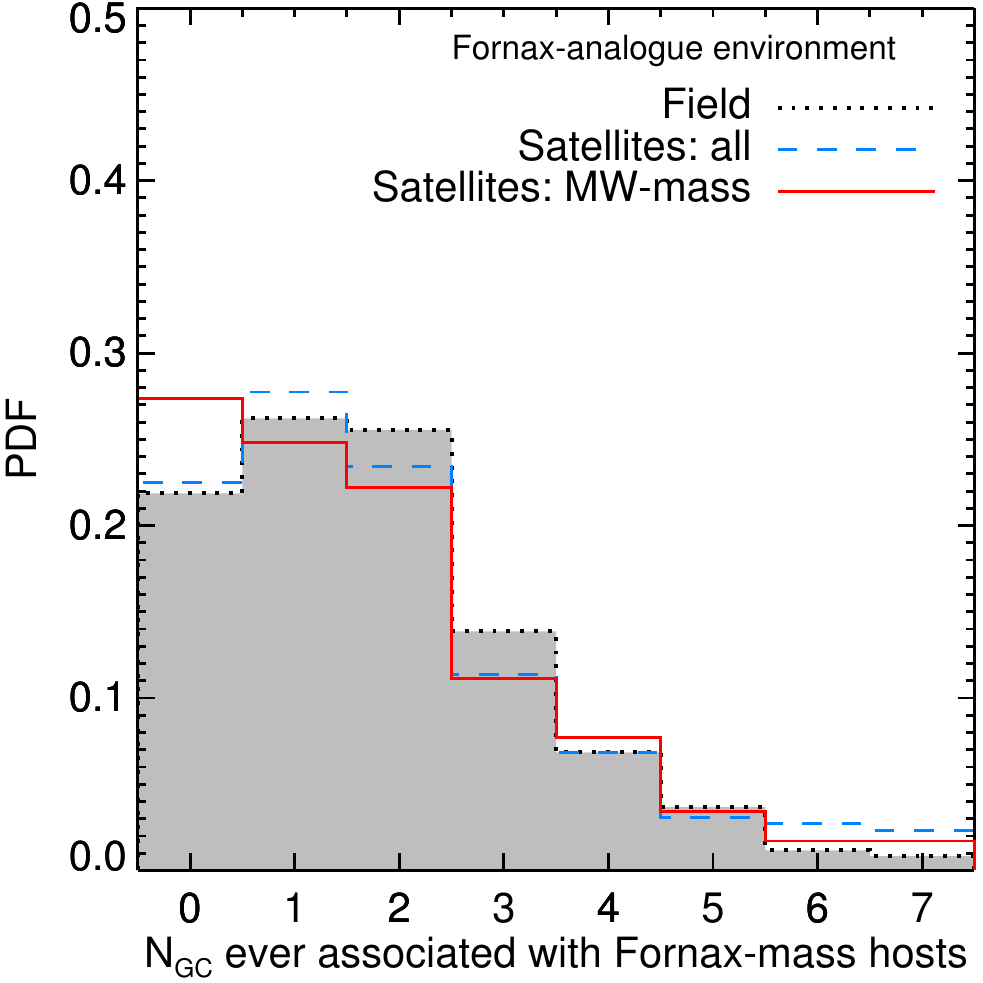}
    \caption{ \textit{Top panel}: PDF of the number of GCs at $z=0$ in
      Fornax analogues. We only show GCs with present-day masses above
      $4\times10^{4}\Msun$, the lowest mass GC in Fornax. The three
      lines show the distribution for Fornax-mass dwarfs in the field
      (dotted line), for satellites around all hosts (dashed line) and
      for satellites around MW-mass hosts (solid
      line). \textit{Bottom}: same as the top panel but counting also
      the GCs more massive than $4\times10^{4}\Msun$ that were tidally
      disrupted after sinking to the centre of their host galaxy. }
	\label{fig:Pdf_num}
\end{figure}
%%%%%%%%%%%%%%%%%%%%%%%%%%%%%%%%%%%%%%%%%%%%%%%%%%%%%%%%%%%%%%%%%%%%%

\subsection{The number of GCs in Fornax-mass dwarfs}

The first question we address is how frequently do the simulations
produce dwarfs with the number of GCs observed in the Fornax dSph,
that is at least five with mass,
$M_{\rm GC}>4\times10^{4}\Msun$?  We answer this question in the
top panel of \reffig{fig:Pdf_num}, which shows the distribution of the
number of GCs that survive to $z=0$ in our Fornax analogue sample.
Regardless of whether they are satellites or field galaxies, nearly
$35~$per cent of the Fornax-mass hosts do not have any GCs at all and $30~$per
cent have only one object. Thus, \emosaic{} predicts a
low number of GCs in Fornax-mass dwarfs. Field dwarfs tend to have
more GCs than satellite galaxies, but the effect is rather small. The
difference could be due to: i) satellite dwarfs having stopped forming
stars, and hence GCs, after falling into their host \citep[although
infall can lead to at least a temporary increase in star formation,
e.g.][]{Shao2018b,Genina2019,Hughes2019}, and ii) having had some of the GCs
that were on extended orbits tidally stripped by their more massive
host. 

When comparing with the Fornax dSph we find that only $3~$per cent of
the simulated analogues have five or more GCs. Thus, systems of GCs as
rich as the one observed in Fornax are predicted to be rather rare. As
we mentioned in the previous section, around $33~$per cent of the GCs
that formed within, or were accreted into, our Fornax analogues sunk
to the centre of these systems and were tidally destroyed. So, could
the rarity of GC-rich dwarfs in our model be due to tidal destruction
of GCs? We answer this question in the bottom panel of
\reffig{fig:Pdf_num} where we again show the PDF of the GC count for
each Fornax-mass dwarf, but now we include also the tidally destroyed
GCs. Thus, the panel shows the \emosaic{} prediction for the PDF of
all the GCs that ever formed within, or fell into, a present-day Fornax-mass
dwarf.

As expected, we find an increase in the number of GCs per dwarf
galaxy. In particular, we find a decrease in the fraction of dwarfs
with none or one GC, and an increase in systems with two or more
GCs. However, the tail of the distribution corresponding to five or
more GCs changes only slightly, with the chance of having a GC system
as rich as the Fornax dSph increasing to only $6.5~$per cent. This value is still rather low and indicates that the tidal destruction of GCs is
not the main factor responsible for the low prevalence in \emosaic{}
of GC systems as rich as the one in Fornax. The main difference between haloes with a central core and those with a cusp is the fraction of tidally disrupted GCs, which is expected to be lower for the former case \cite[e.g.][]{Goerdt2006,Meadows2020}. Thus, the number of Fornax GCs is not a reflection of whether or not the dark matter halo profile has a central core or cusp.

The number of GCs in a galaxy depends  strongly on the mass cut used
for the GC selection. Here, we only select GCs with a present-day
stellar mass above $4\times10^4\Msun{}$, which is the mass of the
lightest GC in Fornax. The uncertainty in the mass estimate of that GC
is ${\sim}1\times10^4\Msun$ (25~per cent fractional error), as shown
in \reftab{tab:Fornax_GC_properties}. To have a fair comparison
between our model prediction and the data we should include this
measurement error in our predictions. We recalculate the fraction of
hosts that have five or more GCs by adding a random error to the GC
mass, given by a Gaussian distribution with zero mean and
standard deviation, $1\times10^4\Msun$. We find that the inclusion
of errors leads to a small increase in the fraction of GC systems as
rich as the one in  Fornax. For %example, for 
our fiducial case that
includes GC tidal disruption, the prevalence of Fornax GC systems
grows from 3~per cent, when no mass error is included, to 4~per cent 
when modelling mass measurement errors. 

The \emosaic{} predictions have been shown to agree well with
observational data \citep{Kruijssen2019a} such as the total mass in
GCs at a given host halo mass or host galaxy stellar mass \citep[for
details see Fig.~1 in][]{Bastian2020}. This suggests that the
paucity of rich Fornax-like GC systems is unlikely to be due to a
failure of the GC formation and evolution model and that, most likely,
\textit{suggests that Fornax has an excess of GCs for its stellar
  mass.}

A census of GCs in galaxies as faint as Fornax is currently lacking
and the available observations are rather heterogeneous, which makes
it difficult to perform a robust comparison between Fornax and other
equal-mass dwarfs. However, from the currently available data in
\citet{Forbes2018}, which provides a table of all dwarf galaxies with
one or more GCs, we conclude that our predictions in
\reffig{fig:Pdf_num} are broadly consistent with observations. For
example, the table of \citeauthor{Forbes2018} contains 9 dwarfs with
luminosity within $\pm 1~\rm{mag}$ of the Fornax dSph. Of those, the
majority (7 out of 9) have either one or two GCs, and only two
systems, Fornax and UGC685, have more than two (interestingly, both
these galaxies have 5 GCs).

%%%%%%%%%%%%%%%%%%%%%%%%%%%%%%%%%%%%%%%%%%%%%%%%%%%%%%%%%%%%%%%%%%%%%
\begin{figure}
	\plotone{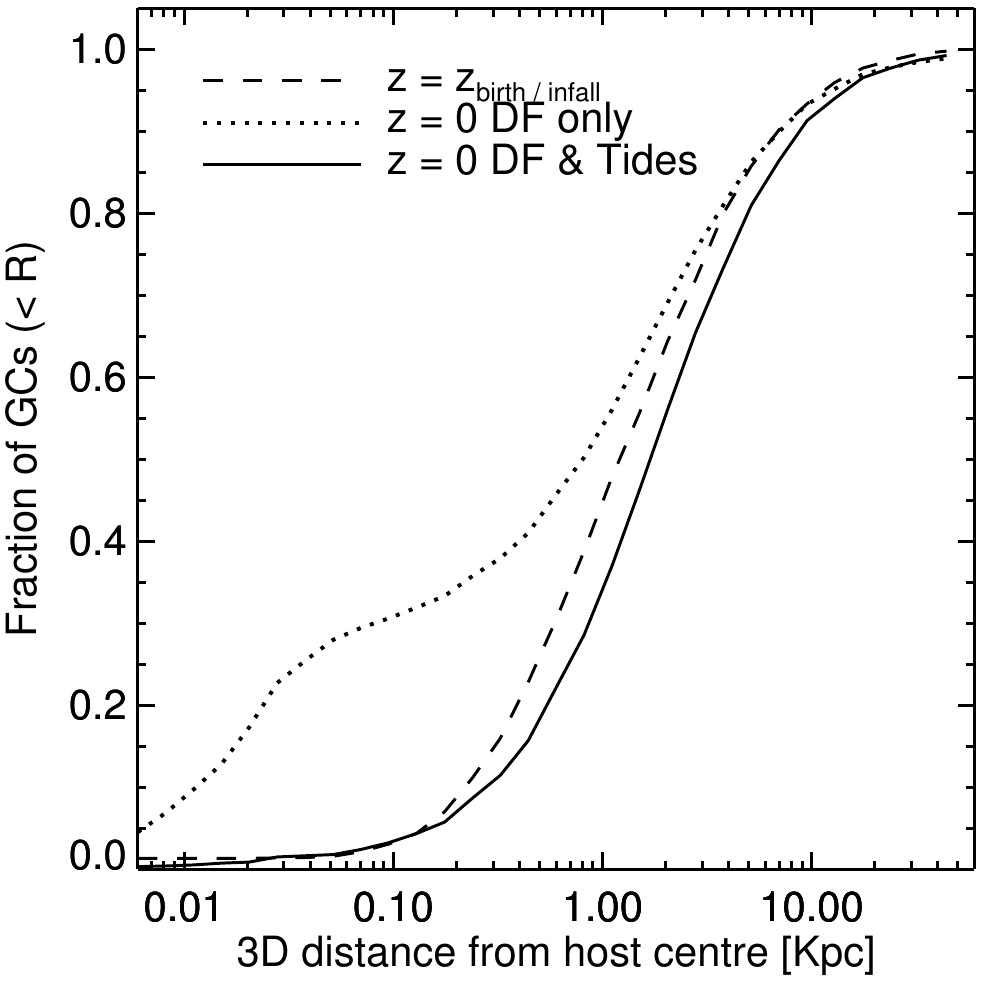}
	\caption{The 3D radial distribution of GCs in Fornax-mass
          dwarfs. The dashed line shows the distribution of GCs at
          birth, if they formed {\em in situ}, or at infall into their
          Fornax analogue host, if they were accreted. Dynamical
          friction leads to some GC sinking to the centre of their
          host and the resulting $z=0$ distribution is shown by
          the dotted line. Once a GC gets close enough to the centre,
          it is disrupted by the tides of the host galaxy. The
          resulting $z=0$ radial distribution is shown by the solid
          line and includes the effect of both dynamical friction and
          tidal disruption.
	}
	\label{fig:Pdf_radii}
\end{figure}
%%%%%%%%%%%%%%%%%%%%%%%%%%%%%%%%%%%%%%%%%%%%%%%%%%%%%%%%%%%%%%%%%%%%%

\subsection{The radial distribution of GCs}

Since, as we have seen, the high number of GCs in Fornax is not in itself a reliable way to infer if the DM halo is cuspy or has a core.
In this
section we examine the constraints that can be placed on the inner
structure of Fornax by considering instead the radial distribution of
its GCs.
Most dwarfs have one or two GCs, so it is difficult and not very
meaningful to show the radial distribution for each host. Instead, we
stack all our Fornax-mass dwarfs and study the {\em mean} radial
profile of GCs obtained by stacking the distances to their host's 
centre.
 
We start with \reffig{fig:Pdf_radii} which shows how the 3D radial
distribution of GCs changes from their initial position to the present
day. The dashed line shows the initial positions of GCs which
correspond to either their birth location (for {\em in situ} GCs) or
to their position at the time they were accreted into their $z=0$ host
(for accreted GCs). There is a large range of initial positions, with
a median value of 1.7\kpc{} and 10 and 90 per centiles of 0.3 and 8.5
\kpc{} respectively. Although not shown, we have checked that the
{\em in situ} population is more concentrated than the accreted
sample.

The subsequent orbital evolution leads to the GCs moving inwards
on average, as illustrated by the dotted line in
\reffig{fig:Pdf_radii}, which shows the $z=0$ positions after including
the effect of dynamical friction but without removing tidally
destroyed GCs. We find that around one third of the GC sample sinks to
the centre of their galaxies, i.e. to $r<0.1\kpc$, and thus are
likely to be tidally disrupted. When accounting for tidal disruption, we find that the majority of GCs
with $r\lesssim0.2\kpc$ are removed and the distribution of surviving
GCs, which is shown by the solid line in \reffig{fig:Pdf_radii}, is
now somewhat less concentrated than the initial GC positions.

%%%%%%%%%%%%%%%%%%%%%%%%%%%%%%%%%%%%%%%%%%%%%%%%%%%%%%%%%%%%%%%%%%%%%
\begin{figure}
	\plotone{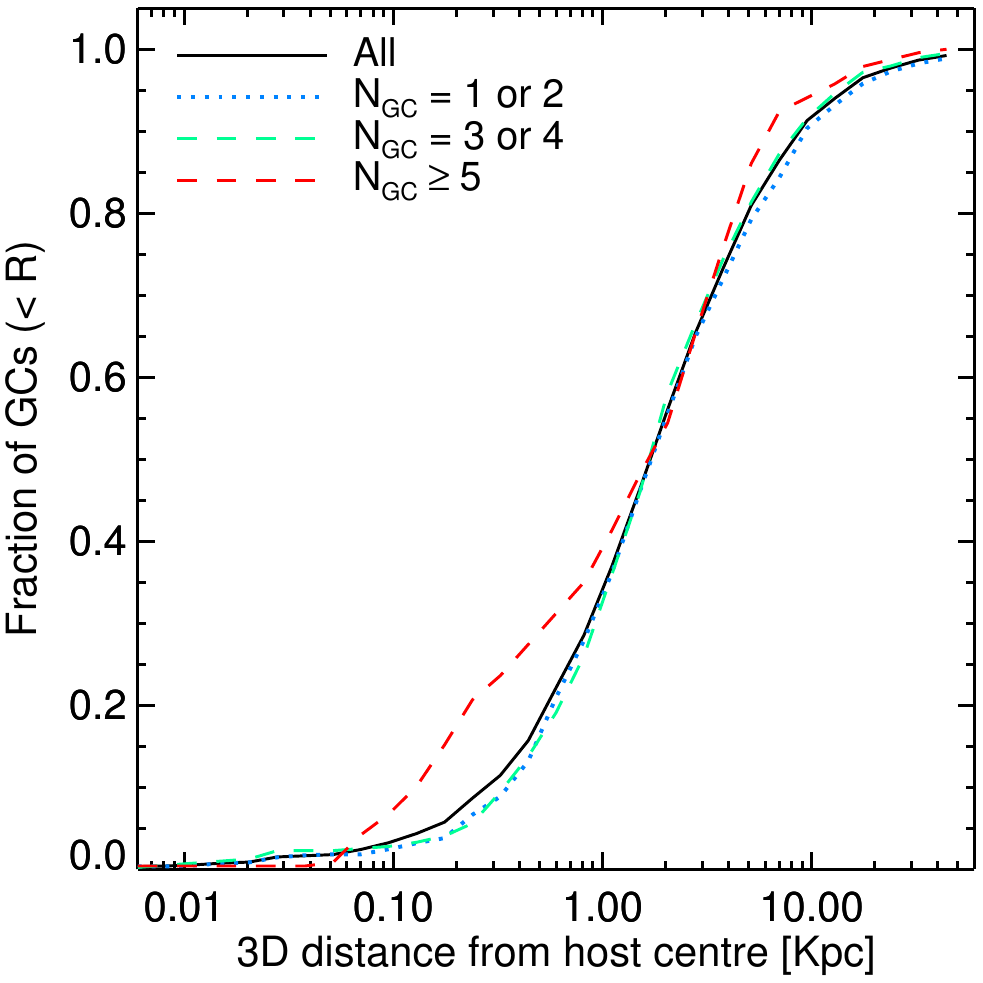}
	\caption{The dependence of the 3D radial distribution of GCs
          on the number of GCs in a Fornax-mass dwarf. It shows that
          there is no correlation between the radial distribution of
          GCs and the number of GCs in a galaxy.}
	\label{fig:Pdf_radii_mul}
\end{figure}
%%%%%%%%%%%%%%%%%%%%%%%%%%%%%%%%%%%%%%%%%%%%%%%%%%%%%%%%%%%%%%%%%%%%%

In \reffig{fig:Pdf_radii_mul} we investigate if GC systems in all
Fornax-mass dwarfs have the same average radial profile. In
particular, we study if the radial profile depends on the number of
GCs in a given galaxy. This test is motivated by our previous result
that Fornax has a large number of GCs for its stellar mass and we wish
to investigate if such an excess also impacts the GC radial
distribution. The figure shows that there is no significant
correlation between the number of GCs and their radial distribution.
(While the dashed line shows some deviations from the mean trend,
these are consistent with random scatter given the small sample size.) In
practice, this shows that no systematic bias is introduced when
comparing the Fornax GC distribution with the average profile of the
full Fornax analogue sample.

%%%%%%%%%%%%%%%%%%%%%%%%%%%%%%%%%%%%%%%%%%%%%%%%%%%%%%%%%%%%%%%%%%%%%
\begin{figure}
	\plotone{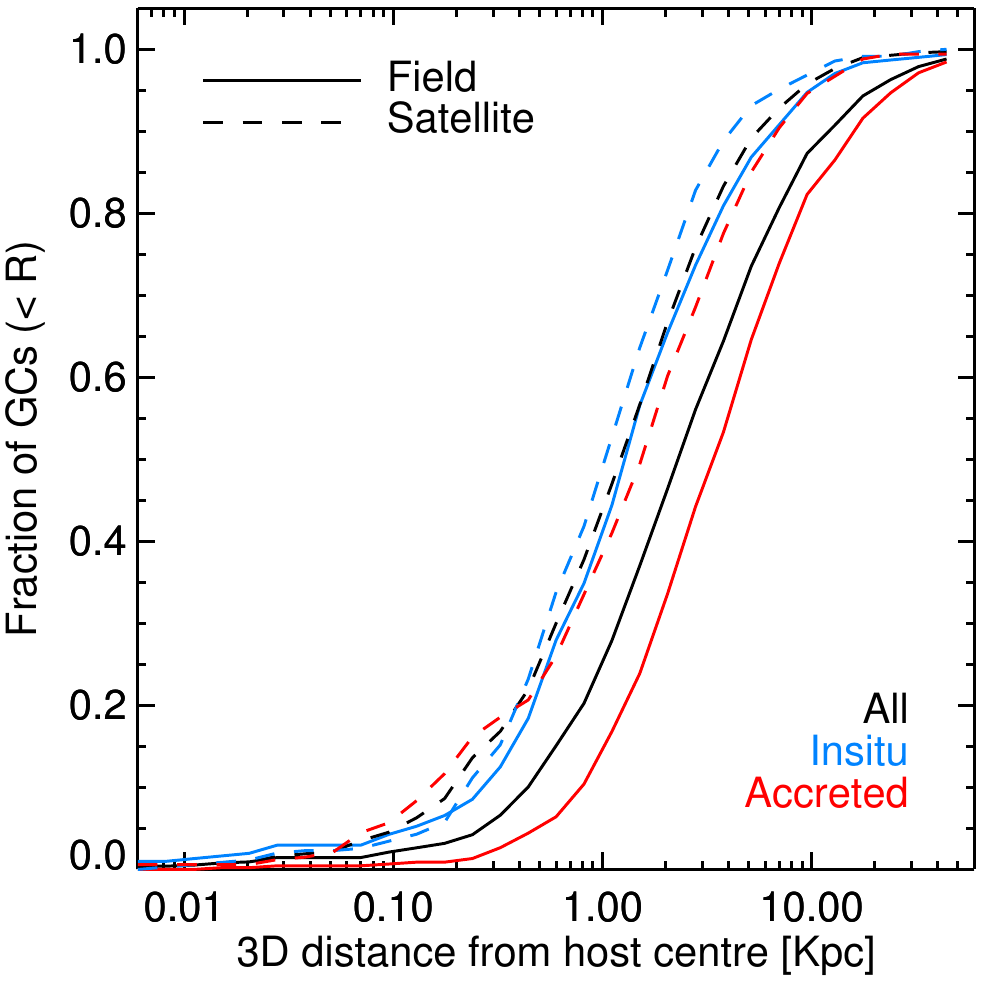}
	\caption{Comparison of the present-day radial distribution of
          GCs in field (solid lines) and satellite (dashed lines)
          Fornax-mass hosts. For each subsample we show the
          distribution of all GCs (in black), those formed {\em in
            situ} (in blue) and those accreted (in red). }
	\label{fig:Pdf_radii_type}
\end{figure}

%%%%%%%%%%%%%%%%%%%%%%%%%%%%%%%%%%%%%%%%%%%%%%%%%%%%%%%%%%%%%%%%%%%%%

\reffig{fig:Pdf_radii_type} investigates how the radial distribution
of GCs differs in satellite and field dwarfs. GCs in satellite dwarfs
(black-dashed line) are more radially concentrated than those in field
dwarfs (black-solid line): nearly half of the GCs in satellites are
located within $1$\kpc{} from the centre while only 25~per cent in the
field dwarfs are. The difference between satellite and field dwarfs is due to
tidal stripping of GCs in satellite dwarfs, which preferentially
removes objects on extended orbits and thus leads to more concentrated
distributions. While not shown, we have checked this by tracing back
the satellite and field dwarfs to redshift, $z=2$, and comparing their
GC distributions at that time. To make a meaningful comparison, for
the $z=0$ satellites, we used only the $z=2$ progenitors that were
central galaxies, which represent the bulk of the population. We find
that at redshift $z=2$ there is no significant difference in the GC
distributions of progenitors of satellite and field dwarfs. Thus,
the differences seen in \reffig{fig:Pdf_radii_type} are mainly due to
tidal stripping from the outskirts of satellite dwarfs.

In \reffig{fig:Pdf_radii_type} we further split the GCs into two
additional subsamples: accreted and {\em in situ}. In both satellite
and field Fornax analogues, the {\em in situ} GCs are more radially
concentrated than the accreted population. The segregation between the
{\em in situ} and accreted components is largest for field dwarfs. The
fraction of accreted GCs in satellite dwarfs is $50~$per cent, lower
that the $60~$per cent fraction in field dwarfs. This difference could
be due to two effects. First, satellite galaxies experience mergers
mostly before infall into a more massive host halo \citep[see
e.g.][]{Angulo2009,Deason2014}; they thus have less time to devour
other galaxies and grow their accreted GC population. Secondly, as we have just seen, accreted GCs have a more radially-extended
distribution than the {\em in situ} GCs and so they are more easily
tidally stripped from their hosts.

%%%%%%%%%%%%%%%%%%%%%%%%%%%%%%%%%%%%%%%%%%%%%%%%%%%%%%%%%%%%%%%%%%%%%
\begin{figure}
	\plotone{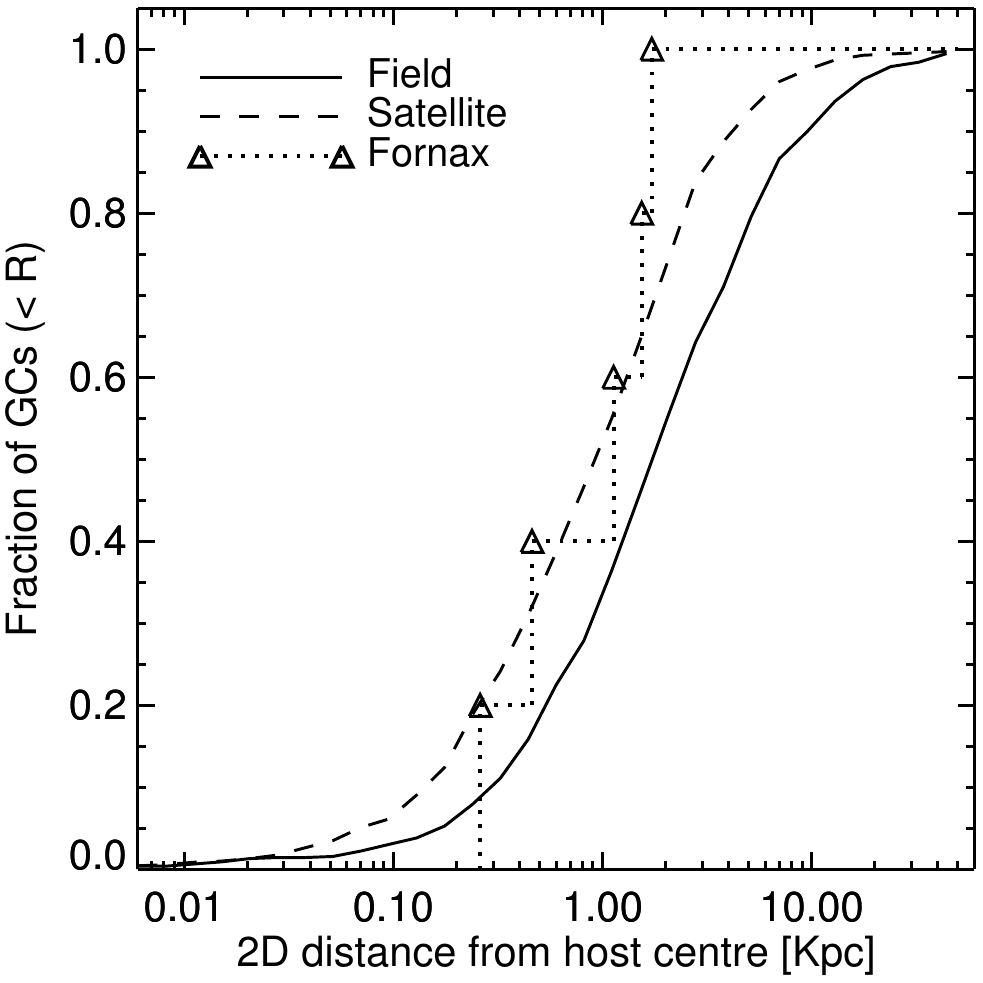}
	\caption{The distribution of projected 2D distances of GCs in
          Fornax analogues. The solid and dashed lines show the
          results for respectively field and satellite Fornax-mass
          dwarfs. The observed projected distances of GCs in Fornax
          are shown by the dotted line (each triangle symbol
          corresponds to a Fornax GC). The distribution of GCs in
          Fornax matches very well that of its analogues in the
          simulation. 
    }
	\label{fig:Pdf_radii_2d}
\end{figure}
%%%%%%%%%%%%%%%%%%%%%%%%%%%%%%%%%%%%%%%%%%%%%%%%%%%%%%%%%%%%%%%%%%%%%

Finally, we compare the GC radial distribution of our Fornax analogues
with that of the real Fornax dSph. We take the projected distances of
Fornax GCs from \citet{Mackey2003}, which we recalculate using a more
recent distance estimate for Fornax of $147\pm4$ kpc
\citep{McConnachie2012}; this new distance is about $7~$per cent
larger than the 137 kpc used by \citeauthor{Mackey2003}. The updated
GC distances are given in \reftab{tab:Fornax_GC_properties} and their
radial distribution is shown in \reffig{fig:Pdf_radii_2d}, where we
also show the projected distance distribution predicted by our model
for satellite and field dwarfs. Qualitatively, we find good
agreement between observations and the simulations. To quantify the
degree of agreement, we use the Kolmogorov-Smirnov (KS) test. The KS
test returns a p-value, $p=0.98$, when comparing the Fornax GCs with
those of satellite analogues in the \emosaic{} simulation, and a lower
value, $p=0.55$, when comparing with the field analogues. This
indicates that any differences seen in \reffig{fig:Pdf_radii_2d} are
consistent with random noise and thus not significant.

\section{Discussion}
\label{sec:discussion}

In this section we discuss in detail the implications of our results
and what they can tell us about the formation history of the Fornax
dSph.

\subsection{The DM halo of the Fornax dSph}
\label{sec:discussion:Fornax_DM_halo}

Our stellar-mass selected Fornax analogues are found in haloes of
present-day mass, $M_{200}=2.8^{+1.1}_{-1.0}\times 10^{10} \Msun$
(68~per cent confidence limit), and concentration,
$c=10.5^{+3.6}_{-2.6}$. These values apply to field galaxies,
i.e.~to Fornax-mass dwarfs that are central galaxies. The $M_{200}$ and
$c$ values can be used to infer the halo scale radius, $r_s$, and
characteristic density, $\rho_0$ (see Eq. \ref{eq:NFW_profile}). The
resulting values of $r_s$ and $\rho_0$ provide a good description also
of the inner region ($r\lesssim r_s$) of our Fornax-mass satellites,
as long as these satellites have not experienced stripping of their
inner DM profile. This is because DM halo growth proceeds primarily by
the deposition of newly accreted material in the outer parts of the
halo with the inner parts remaining largely unchanged
\citep{Wang.Jie2011}. Once a galaxy becomes a satellite, it stops
accreting mass and, in fact, can lose mass by tidal stripping, with
most of the decrease taking place at the outskirts of the halo while
the inner region changes only slowly
\citep[e.g.][]{Penarrubia2008,Errani2020}. The profile of the inner
halo changes appreciably only once a large fraction ($\gtrsim50~$per
cent) of the mass has been lost, and this will be manifest in changes in the
orbits of stars and GCs \citep{Penarrubia2008,Fattahi2016}. Thus,
as long as the satellites have not suffered a large degree of tidal
stripping, we expect that they should have similar inner DM profiles as
their field counterparts (we explicitly show this for the halo scale
radius, $r_s$, in \reffig{fig:Pdf_Fornax}).

We have found that the Fornax dSph has an atypically large number of
GCs for its stellar mass. Observations have shown that the number and,
in particular, the total mass in GCs correlates with halo mass
\citep[e.g.][]{Forbes2018}. The same correlation is well reproduced in
the \emosaic{} simulation \citep{Bastian2020}. This raises the
question of whether we can take into account the richness of the
Fornax GC system to better determine properties of its dark halo. We
investigate this possibility in
Appendix~\ref{appendix:GC_rich_dwarfs}, where we show that our Fornax
analogues that have at least three GCs (the fraction of systems with
at least five GCs is too low to obtain robust conclusions) reside in
DM haloes that are, on average, 20~per cent more massive than the full
Fornax-mass sample. Interestingly, the DM halo concentration does not
show a trend with GC abundance.

Our average, our DM halo mass estimates are an order of magnitude
larger than previous determinations based on the stellar kinematics of
the Fornax dSph, which suggest a DM halo mass of
$\rm{few}\times10^9\Msun$
\citep{Goerdt2006,Boldrini2019,Meadows2020}. However, our estimates
are in good agreement with the recent non-parametric Jeans modelling
of \citet{Read2019}, who found a total mass of
${\sim}2\times10^{10}\Msun$. The discrepancy could be due to the
Fornax halo having experienced a large amount of tidal stripping, as
suggested by \citet{Fattahi2016}, \citet{Wang.M.Y.2017} and
\citet{Genina2020}. These studies used hydrodynamical simulations to
select Fornax counterparts by matching the observed stellar
velocity-dispersion and stellar half-mass radius in the first two of
these papers and the stellar mass in the last one. If stripping is
important then a ${\sim}10^9\Msun$ halo is not representative of the
halo in which the Fornax GCs formed and evolved for most of their
lifetime before the galaxy fell into the Milky Way.

%%%%%%%%%%%%%%%%%%%%%%%%%%%%%%%%%%%%%%%%%%%%%%%%%%%%%%%%%%%%%%%%%%%%%
\begin{figure*}
	\plotone{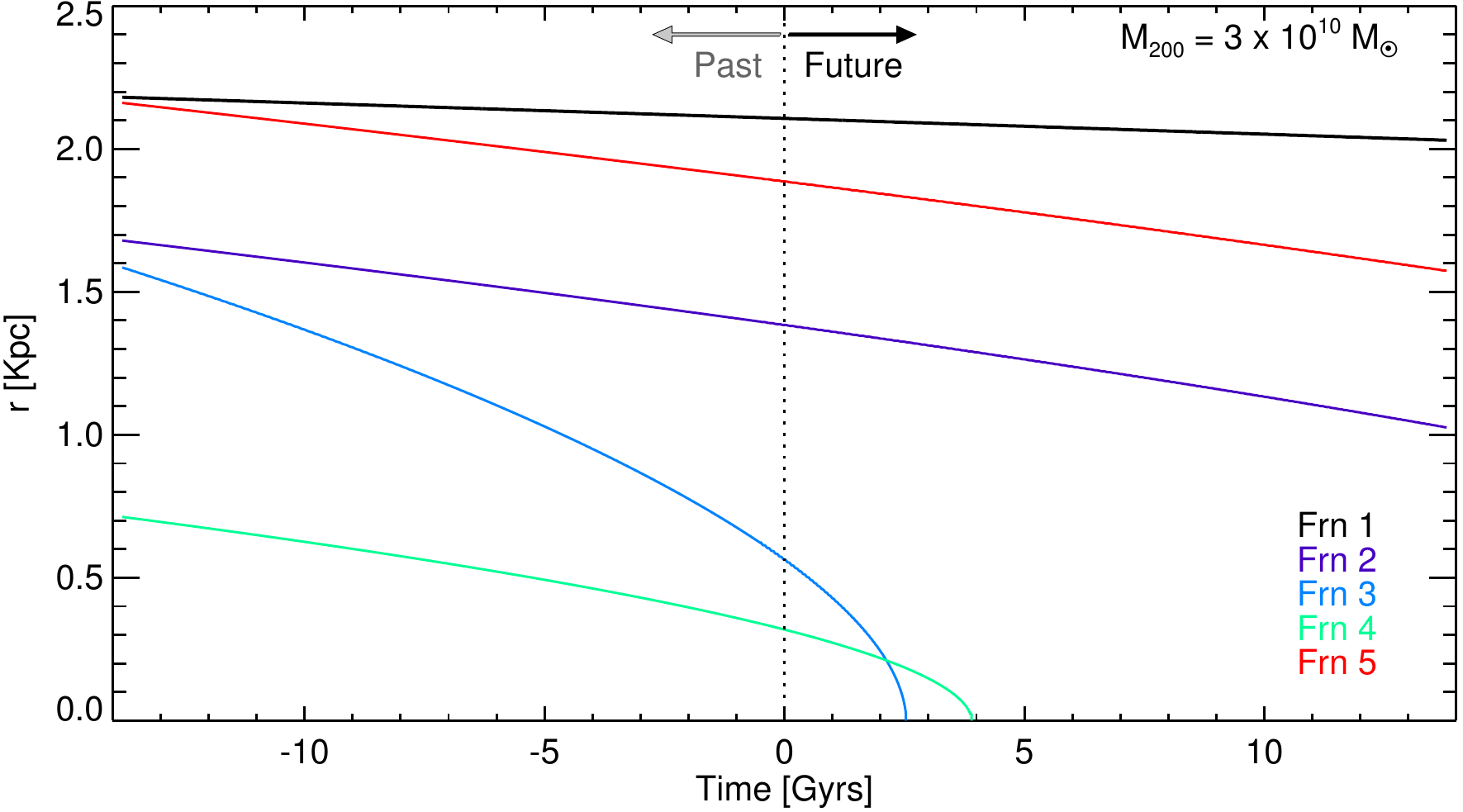}
	\caption{The orbital evolution of the GCs in Fornax assuming the
          median mass, $M_{200}=3\times10^{10}\Msun$, and the median
          concentration, $c=10$, for the DM halo of Fornax's. Negative time
          values correspond to past orbits and positive ones to
          future orbits. Each colour corresponds to one of the 5 GCs
          in Fornax (see bottom-right legend). In the past the 
          GCs were farther from the galaxy centre, especially Fornax 3,
          which is the most massive, and Fornax 4, which today is the
          closest to the centre. 
	}
	\label{fig:future_past}
\end{figure*}
%%%%%%%%%%%%%%%%%%%%%%%%%%%%%%%%%%%%%%%%%%%%%%%%%%%%%%%%%%%%%%%%%%%%%

\subsection{What determines the sinking time scale for GCs?}
\label{sec:discussion:sinking_time}

Our result agrees with earlier work that showed that in a cuspy DM
halo, the orbits of GCs decay because of dynamical friction
\citep[e.g.][]{Goerdt2006,Meadows2020}. However, we estimate a longer
decay time for the Fornax GCs that in previous work. The differences
arise from two novel aspects introduced into this analysis. Firstly,
we show that previous measurements of the Fornax DM halo based on
simple stellar kinematic analyses give a mass that is too low (see
discussion in \refsec{sec:discussion:Fornax_DM_halo}); this leads to
an estimate of the sinking timescale that is correspondingly too low
(see Section~ \ref{fig:GC_orbit_M200}). On average, the sinking
timescale doubles in an $M_{200}=3\times10^{10}\Msun$ halo compared to
that in a halo of the previously assumed value of
$M_{200}=2\times10^{9}\Msun$.

Secondly, as pointed out in previous work, we know only the
present-day positions of the GCs, not their birth positions
\citep[e.g.][]{Angus2009,Boldrini2019,Meadows2020}. The sinking time
increases rapidly with increasing distance, and even small changes in
distance can have a considerable effect. To take this into account in
our analysis, we used the GC positions and velocities at birth from
the \emosaic{} simulation which self-consistently models the formation
and evolution of GCs in cuspy DM haloes. Our calculation shows that
the majority (67~per cent) of GCs in Fornax-mass dwarfs survive to the
present day and thus only a third are expected to sink to the centre
of their galaxy in a Hubble time. The survival chance depends on the
starting position of the GC: it is as low as 37~per cent for GCs
formed within 1\kpc{} from the centre and as large as 92~per cent for
GCs formed beyond 1\kpc{} from the centre.

\subsection{Past and future orbits of the GCs in
  Fornax} \label{sec:discussion:future_orbit}

Given the updated GC sinking times discussed in
\refsec{sec:discussion:sinking_time}, we recalculate, in
\reffig{fig:future_past}, the possible orbits of the Fornax GCs. The
figure shows both the past (to the left of the vertical dotted line)
and the future orbits assuming the median halo mass, $M_{\rm
  200}=3\times10^{10}\Msun$, and the median concentration, $c=10$, of our
Fornax analogue sample. The distances of GCs are in 3D and the orbit
integration was done assuming circular orbits with present-day radii
of $\sqrt{3/2}$ times their observed projected radial distance. 

This simple model suggests that only one GC, Frn~4, was born at a
small radial distance, $r_{\rm init}=700\pc$, which is approximately
the half-light radius of the present-day Fornax (as given by \citealt{McConnachie2012}). For the remaining
four GCs, the initial distances are larger than $1.5$\kpc{}, which
suggests that at least some of them were accreted \citep[see
also][]{Boldrini2020}. This result is in good agreement with our model prediction
that around half of the present-day surviving GCs are of accreted
origin. The sinking time varies from GC to GC depending on the initial
distances and masses; Frn 1, Frn 2, and Frn 5 could still survive for
another Hubble time mainly because of their large initial
distances. Interestingly, the time for Frn 3 to sink is the shortest
amongst the five even though it started from $r_{\rm init}=1.5\kpc$,
which is twice as far as Frn 4. This is due to the fact that Frn 3 is
the most massive GC of the five, with a mass, 
$M_\star=5\times10^5\Msun$.

We note that the orbital evolution in \reffig{fig:future_past}
is highly simplified and does not take into account that the GCs are
likely to be on eccentric orbits. For example, \emosaic{} predicts
that at birth the GCs in Fornax-mass dwarfs have an orbital
ellipticity of $0.64^{+0.20}_{-0.28}$ (68th per centile). Furthermore,
there is indirect evidence that the Fornax dSph might have had a major
merger ${\sim}10$ Gyrs ago \citep{Yozin2012,Benitez-Llambay2016} and
that it is potentially undergoing severe tidal stripping
\citep{Wang.M.Y.2017,Genina2020}. The latter process, if it takes
place rapidly enough, could lead to the dissolution of the GC system in the outskirts of the Milky Way halo
before Frn 3 can sink to the centre.

\subsection{GCs and the core-cusp problem} \label{sec:discussion:core_vs_cusp}

The existence of GCs close to the centre of Fornax has often been
adduced as evidence that this galaxy's DM halo has a constant density
core. Here we have shown, to the contrary, that cuspy DM haloes with
galaxies of simliar stellar mass to Fornax are expected to have a
similar GC radial distribution: the GCs in Fornax cannot be used to
rule out a cuspy halo \citep[see also][]{Angus2009,Meadows2020}. In
fact, the absence of an extended core, of radius $\sim 1\kpc$, in
Fornax is supported by other probes. \citet{Read2019} used
higher-order moments of the velocity distribution to break the
degeneracy between mass and velocity anisotropy and found that the
stellar kinematics in Fornax favour an increase of an order of
magnitude in DM density between radii of 1 and 0.1 \kpc{}, with a core,
if present, having a radius $\lesssim 0.2\kpc$. Such a core radius is
too small to lead to a stalling of the orbital decay of GCs, which
takes place at around ${\sim}1/3$ of the core radius
\citep{Meadows2020}.

Our model predicts that a third of the GCs associated with Fornax-mass
dwarfs have sunk to the centre of their host galaxies where they may
have been tidally disrupted. It remains to be seen what is the
corresponding fraction of destroyed GCs if the DM profile had a core,
but it is likely lower. Thus, identifying in observations the
fraction of GCs that were destroyed is, in principle, a promising
avenue for distinguishing between a cusp and a core in Fornax. The
extent of tidal stripping experienced by GC is sensitive to the
central DM density profile: in cuspy haloes GCs are more extended and
have lower mass loss that in haloes with cores
\citep[e.g.][]{Amorisco2017,Webb2018,Orkney2019} but it is unclear if
this difference can be used as a test of the inner structure of the
halo.

\section{Conclusions}
\label{sect:conclusions}

We have investigated the number and radial distribution of GCs in the
Fornax dSph with the goal of testing whether the GC population in this
galaxy is consistent with a cuspy halo, the simplest prediction of the
$\Lambda$CDM cosmological model \citep{Navarro1996,Navarro1997}. Our
study has been motivated in part by the ongoing debate on whether the
Fornax GCs are a robust signature of a kiloparsec-sized core at the
centre of this galaxy's halo \citep[e.g. see discussion
in][]{Meadows2020}. To this end we have analyzed the \emosaic{} simulation
\citep{Pfeffer2018,Kruijssen2019a}, which includes a
subgrid prescription for self-consistently following galaxy and GC
formation and evolution within the \eagle{} cosmological hydrodynamics
framework \citep{Schaye2015,Crain2015}.

We have proceeded by identifying a sample of Fornax analogues,
selected to have a similar stellar mass, within a factor of two, as the
Fornax dSph, i.e.~in the range $[2, 8]\times10^7\Msun$. \emosaic{} is
well suited for this study because it contains a large sample (1154)
of Fornax-mass dwarfs, within which GC formation and evolution is
followed in a full cosmological context. To account for the GC orbital
decay due to dynamical friction, we have post-processed the orbits of
GCs starting from their initial positions as determined by \emosaic{},
and assuming that they evolve in the NFW profile that best fits each
host halo in the simulation. The initial conditions are given by the
GC positions and velocities at birth for {\em in situ} GCs, and at
infall for accreted GCs. We applied a tides model to account for the
disruption of GC by their host halo. We then investigated the number
and radial distribution of GCs of stellar mass,
$M_\star\geq4\times10^4\Msun$, corresponding to the smallest of the
five well-studied GCs in Fornax. Our main conclusions may be
summarised as follows:
\\[-.55cm]
\begin{enumerate}
 	\item Field Fornax-mass dwarfs reside in haloes of median
          mass, $M_{\rm 200}\approx2.8\times10^{10}\Msun$, and
          concentration, $c\approx10$ (see
          \reffig{fig:Fornax-sample}). 
 	%Rich GC systems reside in more massive haloes with a median $M_{\rm 200}=3.4\times10^{10}\Msun$ for dwarfs that have three or more GCs (see \reffig{fig:Pdf_m200_con_new}).
%  	\\[-.3cm]
%  	\item The DM radial density of dwarfs follow NFW profile with concentration $c\sim10$ and characteristic radius $R_s\sim6\kpc$, with field dwarfs having slight larger $R_s$ than satellite dwarfs due to tidal disruption (see \reffig{fig:Fornax-sample}).
 	\\[-.3cm]
      \item The population of GCs of Fornax analogues consists, in
        nearly equal amounts, of {\em in situ} (45~per cent) and
        accreted objects (55~per cent).
 	\\[-.3cm]
      \item The GCs of Fornax are typically old, with a median age,
        $t_{\rm age}\approx12$ Gyrs. The fraction of GCs younger than
        $12$ Gyrs is 77~per cent for {\em in situ} GCs and 23~per cent
        for accreted GCs. This is because {\em in situ} GCs, which
        form close to their host centre, are more likely to be
        destroyed, while the accreted GCs started orbital decay at
        larger initial distances (see \reffig{fig:Pdf_GC_tform}).
 	\\[-.3cm]
      \item Orbital decay leads to 33~per cent of GCs sinking to the
        centre where they can be tidally disrupted in a cuspy halo
        (see \reffig{fig:Pdf_radii}).
 	\\[-.3cm]
      \item Most of the Fornax-mass dwarfs have one ($\approx$ 30~per
        cent) or no ($\approx$ 35~per cent) GCs, with very few having
        several. The results are similar for both satellite and field
        dwarfs (see \reffig{fig:Pdf_num}).
 	\\[-.3cm]
      \item The median radial distance of surviving GCs is $1.7\kpc$,
        with 10 and 90 percentiles of $0.3$ and $8.5\kpc$
        respectively (see \reffig{fig:Pdf_radii}). 
 	\\[-.3cm]
 	\item Satellite galaxies have more concentrated GC
          distributions than their field analogues. {\em In situ} GCs
          are also more radially concentrated than their accreted
          counterparts (see \reffig{fig:Pdf_radii_type}). 
\end{enumerate}

Our model predicts that only a small fraction (3~per cent) of
Fornax-mass dwarfs have at least as many GCs as are observed in Fornax.
This result does not pose a challenge to our model since observations
similarly show that Fornax has a surprisingly large number of GCs for
its stellar mass (e.g. see Figure C1 in \citealt{Forbes2018}). In
fact, most observed dwarfs with similar masses to Fornax have, at
most, only one or two GCs, in agreement with our model predictions.

The observed radial distribution of the GCs in Fornax agrees very well
with our model prediction for satellite Fornax analogues (KS test
p-value of 0.98). This indicates that, contrary to previous claims,
the distribution of GCs in Fornax can be reproduced in a DM halo that
has a central cusp. Furthermore, our model predicts that the $z=0$
surviving GCs represent ${\approx}$ 67~per cent of the GCs ever associated
with Fornax, ie. born within, or accreted onto, Fornax and thus 
suggests that Fornax may have had an additional ${\sim}2$ GCs that
sunk to its centre and could have been destroyed. One such candidate
has recently been discovered: a low mass GC ${\sim}30$ pc away in
projection from the centre of Fornax which has an irregular shape that
could indicate that it is undergoing tidal disruption
\citep{Wang.M.Y.2019}.

By modelling GC formation and evolution in a cuspy DM halo, our work
has demonstrated that the Fornax GC system cannot be used to rule out
a central cusp in this galaxy's halo. The next steps involve
investigating if the number and radial distribution of GCs differs
between DM haloes with cusps and cores. One way to do so would be to
run \emosaic{} with an increased star formation gas density threshold,
which has been shown to lead to the formation of cores within the
\eagle{} galaxy formation model \citep{Benitez-Llambay2019}. Such a
simulation would predict the GC distribution for profiles with cores
and, by comparing directly against our results, would help identify
which GC statistics are best suited to distinguish between a central
core or cusp in the DM profile.

\vspace{-.2cm}
\section*{Acknowledgements}
%We thank the anonymous referee for detailed comments that have helped us improve the paper.
SS and CSF were supported by the European Research Council through ERC
Advanced Investigator grant, DMIDAS [GA 786910], to CSF and by the
Science and Technology Facilities Council (STFC) [grant number
ST/F001166/1, ST/I00162X/1, ST/P000541/1]. MC acknowledges support by
the EU Horizon 2020 research and innovation programme under a Marie
Sk{\l}odowska-Curie grant agreement 794474 (DancingGalaxies). MRC acknowledges financial support from a CITA National Fellowship. AD and
RAC are supported by Royal Society University Research Fellowships. JP
gratefully acknowledges funding from a European Research Council
consolidator grant (ERC-CoG-646928-Multi-Pop). MRC and JMDK gratefully
acknowledge funding from the European Research Council (ERC) under the
European Union's Horizon 2020 research and innovation programme via
the ERC Starting Grant MUSTANG (grant agreement number 714907). JMDK
gratefully acknowledges funding from the Deutsche
Forschungsgemeinschaft (DFG, German Research Foundation) through an
Emmy Noether Research Group (grant number KR4801/1-1) and the DFG
Sachbeihilfe (grant number KR4801/2-1). This work used the DiRAC Data
Centric system at Durham University, operated by ICC on behalf of the
STFC DiRAC HPC Facility (www.dirac.ac.uk). This equipment was funded
by BIS National E-infrastructure capital grant ST/K00042X/1, STFC
capital grant ST/H008519/1, and STFC DiRAC Operations grant
ST/K003267/1 and Durham University. DiRAC is part of the National
E-Infrastructure.

\section*{Data availability}
%The \eagle{} data are publicly available at \url{http://icc.dur.ac.uk/Eagle/database.php}. 
The used and data produced in this paper are available upon reasonable request to the corresponding author.

\vspace{-.2cm}
\bibliographystyle{mnras}
\bibliography{bibliography}

\appendix

\section{The DM density profile of Fornax-mass dwarfs}
\label{appendix:density_profile}
%%%%%%%%%%%%%%%%%%%%%%%%%%%%%%%%%%%%%%%%%%%%%%%%%%%%%%%%%%%%%%%%%%%%%
\begin{figure}
	\plotone{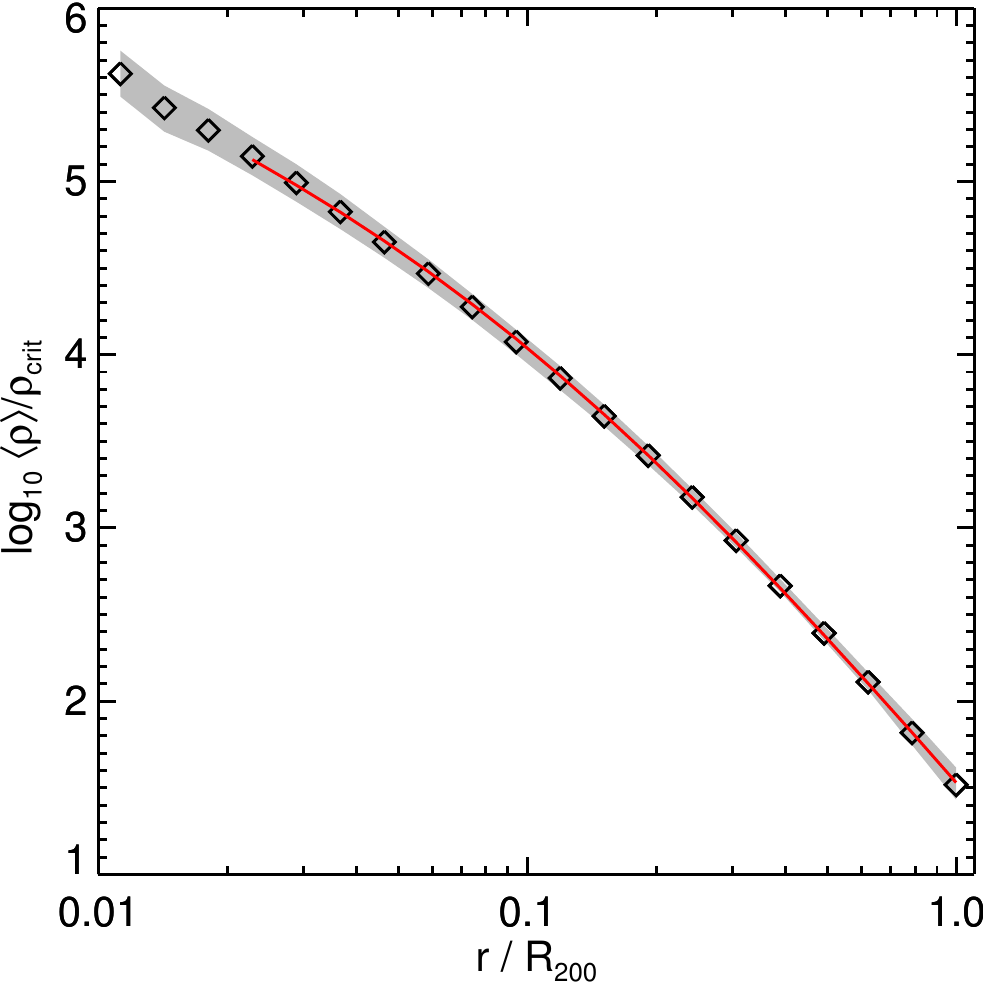}
	\caption{Mean, radially averaged density profile, $\rho(r)$, for the field Fornax-mass haloes in the simulation. The red line shows the best-fit NFW profile.
	}
	\label{fig:rho_all}
\end{figure}
%%%%%%%%%%%%%%%%%%%%%%%%%%%%%%%%%%%%%%%%%%%%%%%%%%%%%%%%%%%%%%%%%%%%%
In agreement with previous results from Eagle \citep{Schaller2015a}, the spherically averaged density profiles of our Fornax-mass haloes are well fit by the NFW profile. This is shown in 
\reffig{fig:rho_all}.

\section{The halo properties of GC rich dwarfs}
\label{appendix:GC_rich_dwarfs}
%%%%%%%%%%%%%%%%%%%%%%%%%%%%%%%%%%%%%%%%%%%%%%%%%%%%%%%%%%%%%%%%%%%%%
\begin{figure}
	\plotone{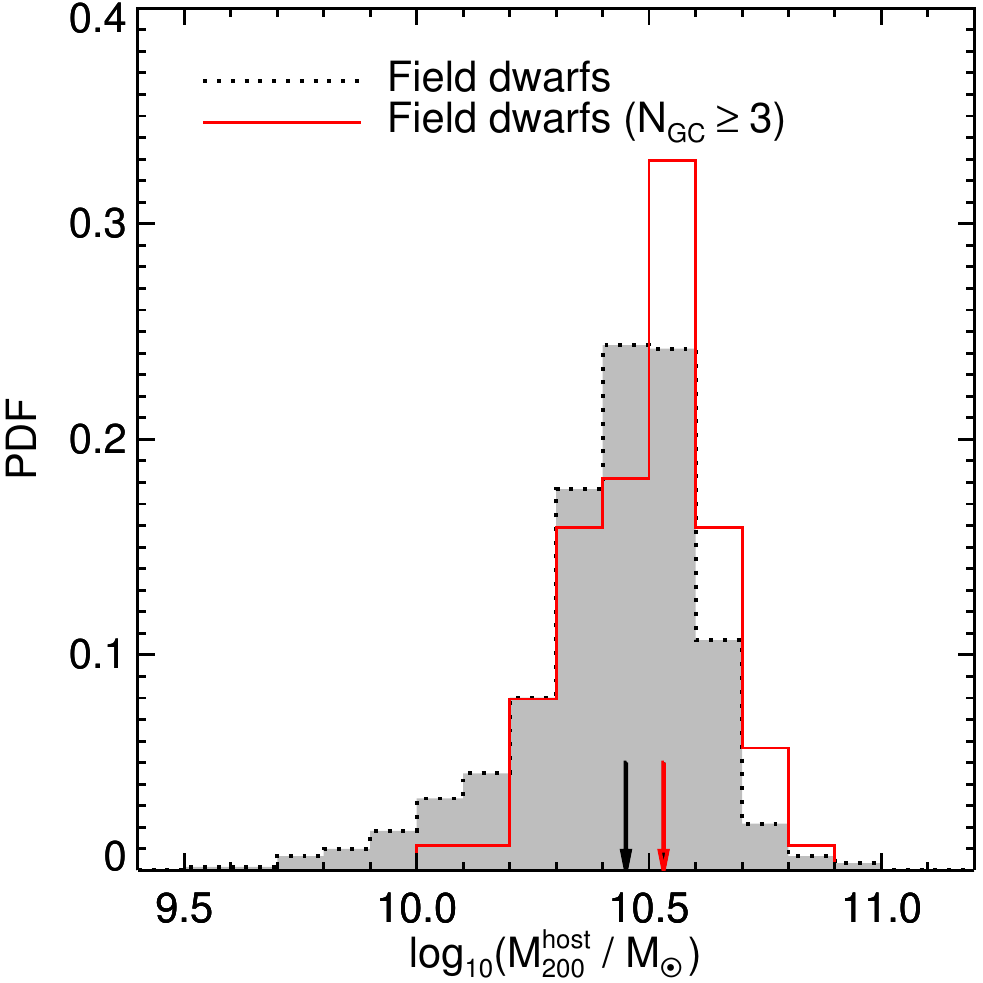} 
	\vskip .2cm
	\plotone{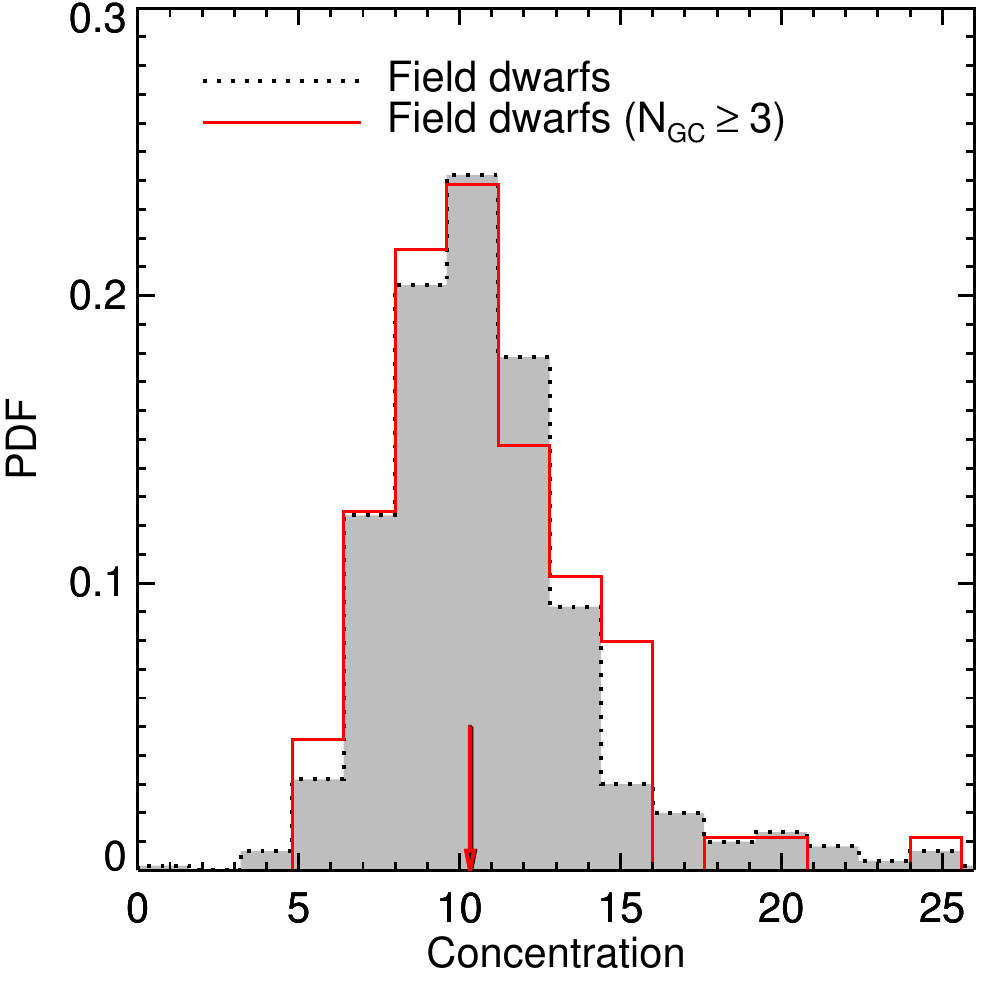}
	\caption{ Comparison of the mass and concentration of DM haloes
          of two Fornax-analogue populations. The black dotted line
          shows the distribution for our reference population of all
          Fornax-mass dwarfs. The red solid line shows the subset of
          systems that have 3 or more GCs. The
          vertical arrows indicate the median of each PDF. 
	}
	\label{fig:Pdf_m200_con_new}
\end{figure}
%%%%%%%%%%%%%%%%%%%%%%%%%%%%%%%%%%%%%%%%%%%%%%%%%%%%%%%%%%%%%%%%%%%%%

In \reffig{fig:Pdf_m200_con_new} we show how the mass and
concentration of the field Fornax analogues varies when, in addition
to selecting the analogues according to stellar mass, we also require
that they have a large number of GCs. To obtain a reasonable sample of
dwarfs with many GCs, we select the subsample that has three or more
GCs. The left-hand panel of \reffig{fig:Pdf_m200_con_new} shows the
PDF of $M_{\rm 200}$ for all the field dwarfs (dotted line) that were
part of our initial, stellar-mass only selection and field dwarfs that
have three or more GCs (solid line). The latter reside in slightly
more massive haloes, with a median,
$M_{\rm 200}=3.4\times10^{10}\Msun$, that is 1.2 times larger than the
median of the full Fornax-analogue sample. This agrees with the fact
that more massive galaxies host more GCs \citep[e.g.][]{Harris2017}. The difference in the
concentrations between the two samples is rather small, as shown in
the right-hand panel.

\section{The mass evolution of GCs in Fornax-mass dwarfs}
\label{appendix:GC_mass_evolution}
%%%%%%%%%%%%%%%%%%%%%%%%%%%%%%%%%%%%%%%%%%%%%%%%%%%%%%%%%%%%%%%%%%%%%
\begin{figure}
	\plotone{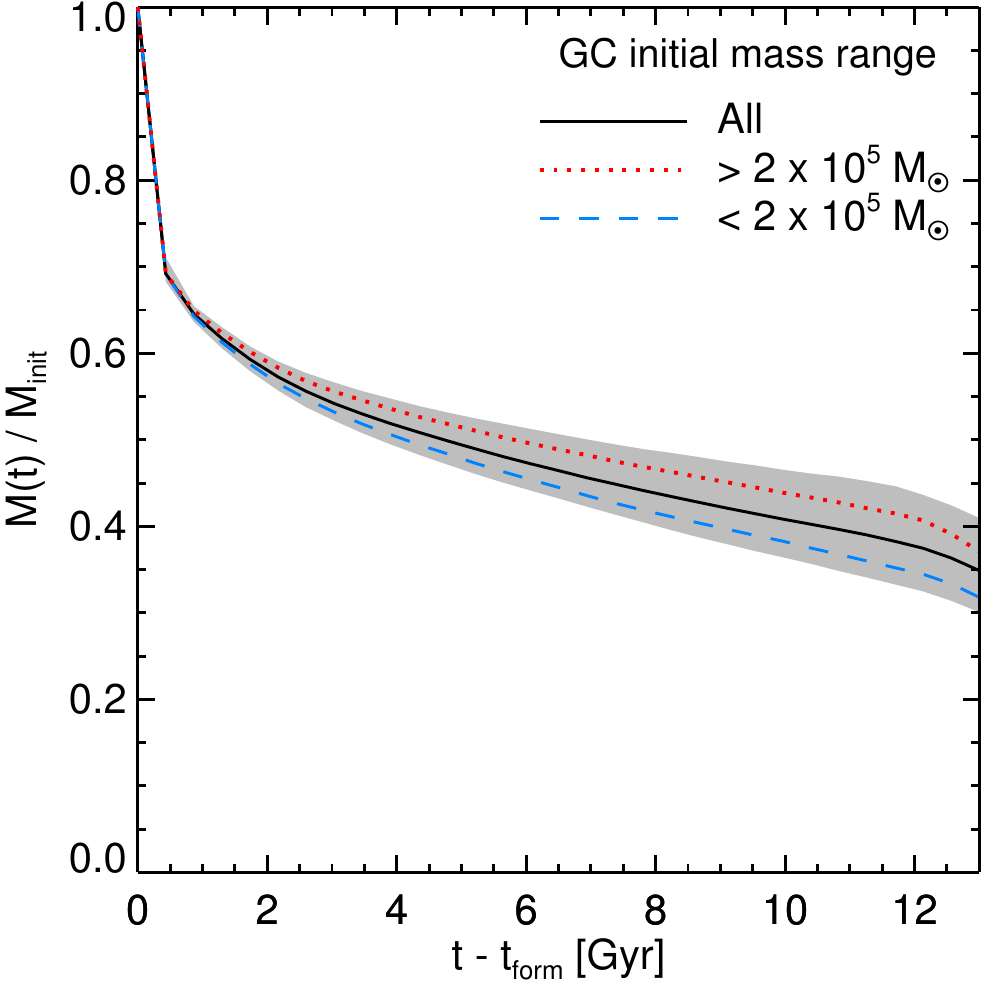}
	\caption{The average mass evolution of our GC sample. The plot
          shows the ratio of the mass at a given time to the initial
          GC mass as a function of the time since the GC formed. The
          result for all the GCs is shown by the black solid line with
          the shaded region showing the 68 percentile
          object-to-object scatter. The various coloured lines show
          the results for GCs in different initial mass ranges (see
          legend). Each of the GC subsamples contains roughly half
          of the full sample. }
	\label{fig:GC_mass_evo}
\end{figure}
%%%%%%%%%%%%%%%%%%%%%%%%%%%%%%%%%%%%%%%%%%%%%%%%%%%%%%%%%%%%%%%%%%%%%

In \refsec{sec:GC_orbit_evolution} we discussed the effect of GC mass
on the decay of its orbit. To make our predictions as realistic as
possible, we accounted for the decrease in GC mass as a function of
time. The average trend, quantified as the ratio of the GC mass at
time $t$ and the initial mass, as a function of the GC age, obtained
from \emosaic{}, is shown in \reffig{fig:GC_mass_evo}. As described in
\refsec{sect:simul}, \emosaic{} accounts for the GC mass lost by 
stellar evolution, two-body relaxation and tidal shocks. When
calculating dynamical friction in post-processing, we assumed that the
GC mass follows the relation shown in the figure (solid black
line). The mass evolution is slightly affected by the initial mass of
the GCs but, for simplicity, we neglect this effect.

\label{lastpage}
\end{document}